\documentclass[usenatbib]{mn2e}
\usepackage{graphicx}
\usepackage{astrobib_mnras2e}

\begin{document}
\topmargin-1cm

\def\bi#1{\hbox{\boldmath{$#1$}}}

\newcommand{\be}{\begin{equation}}
\newcommand{\ee}{\end{equation}}
\newcommand{\bea}{\begin{eqnarray}}
\newcommand{\eea}{\end{eqnarray}}

\newcommand{\lexp}{\mathop{\langle}}
\newcommand{\rexp}{\mathop{\rangle}}
\newcommand{\rexpc}{\mathop{\rangle_c}}

\def\bi#1{\hbox{\boldmath{$#1$}}}
\def\photoz{ photometric redshift }
\def\dndzp{$[dN/dz]_{P}\,$}
\def\dndz{$dN/dz\,$}
\def\erf{{\rm erf}}
\def\sgn{{\rm sgn}}

\title{Calibrating Photometric Redshifts of Luminous Red Galaxies}

\author[Padmanabhan et al]
{Nikhil Padmanabhan$^{1}$\thanks{npadmana@princeton.edu}, 
Tam\'as Budav\'ari$^{2}$,
David J. Schlegel$^{3}$, 
Terry Bridges$^{4}$, \and
Jonathan Brinkmann$^{5}$, 
Russell Cannon$^{6}$,
Andrew J. Connolly$^{7}$,
Scott M. Croom$^{6}$,\and
Istv\'an Csabai$^{8}$,
Michael Drinkwater$^{9}$,
Daniel J. Eisenstein$^{10}$, 
Paul C. Hewett$^{11}$,\and
Jon Loveday$^{12}$, 
Robert C. Nichol$^{13}$, 
Kevin A. Pimbblet$^{9}$, 
Roberto De Propris$^{14}$,\and 
Donald P. Schneider$^{15}$, 
Ryan Scranton$^{7}$, 
Uro\v{s} Seljak$^{1}$, 
Tom Shanks$^{16}$,
Istv\'an Szapudi$^{17}$,\and
Alexander S. Szalay$^{2}$,
David Wake$^{18}$
\\
$^{1}$Joseph Henry Laboratories, Jadwin Hall, Princeton University, Princeton, NJ 08544 \\
$^{2}$Center for Astrophysical Sciences, Department of Physics \& Astronomy, Johns
Hopkins University, Baltimore, MD 21218 \\
$^{3}$Dept. of Astrophysical Sciences, Peyton Hall, Princeton University, Princeton, NJ 08544 \\ 
$^{4}$Physics Department, Queen's University, Kingston, Ontario, Canada, K7M 3N6 \\
$^{5}$Apache Point Observatory, 2001 Apache Point Road, Sunspot, New Mexico 88349-0059 \\
$^{6}$Anglo-Australian Observatory, PO Box 296, Epping,NSW 2121, Australia \\
$^{7}$Department of Physics and Astronomy, University of Pittsburgh, 3941 O'Hara
St., Pittsburgh, PA 15260 \\
$^{8}$Department of Physics, E\"{o}tv\"{o}s University, Budapest, Pf.\ 32,
Hungary, H-1518\\
$^{9}$Department of Physics, University of Queensland, QLD 4072, Australia \\
$^{10}$Steward Observatory, 933 N. Cherry Ave, Tucson, AZ 85721 \\
$^{11}$Institute of Astronomy, Madingley Road, Cambridge CB3 0HA, UK \\
$^{12}$Astronomy Centre,University of Sussex, Falmer, Brighton BN1 9QJ, UK \\
$^{13}$Institute of Cosmology and Gravitation, Univ. of Portsmouth, Portsmouth, PO1 2EG, UK \\
$^{14}$Research School of Astronomy and Astrophysics, Australian National University, 
Weston, ACT, 2611, Australia \\
$^{15}$Department of Astronomy and Astrophysics, Pennsylvania State University, 
University Park, PA 16802 \\
$^{16}$Department of Physics, University of Durham, South Road, Durham 
DH1 3LE, UK \\
$^{17}$Institute for Astronomy, 2680 Woodlawn Road, Honolulu, HI 96822 \\
$^{18}$Dept. of Physics, Carnegie Mellon University, 5000 Forbes Av., 
Pittsburgh,PA 15213 \\
}

\maketitle

\begin{abstract}
We discuss the construction of a photometric redshift catalogue 
of Luminous Red Galaxies (LRGs) from the Sloan Digital Sky 
Survey (SDSS), emphasizing the principal steps necessary for constructing 
such a catalogue -- (i) photometrically selecting the 
sample, (ii) measuring photometric redshifts and their
error distributions, (iii) and estimating the true redshift distribution.
We compare two photometric
redshift algorithms for these data and find that they give
comparable results. Calibrating against the SDSS and SDSS-2dF
spectroscopic surveys, we find that the photometric redshift
accuracy is $\sigma \sim 0.03$ for redshifts less than 0.55 and worsens
at higher redshift ($\sim 0.06$ for $z < 0.7$). These errors are caused by photometric
scatter, as well as systematic errors in the templates, filter curves,
and photometric zeropoints.
We also parametrize the photometric redshift
error distribution with a sum of Gaussians, and use this model to deconvolve
the errors from the measured photometric redshift distribution to
estimate the true redshift distribution. We pay special attention to
the stability of this deconvolution, regularizing the method with a
prior on the smoothness of the true redshift distribution.
The methods we develop are applicable to general photometric 
redshift surveys.
\end{abstract}



\section{Introduction}

Since their inception, photometric redshifts 
\citep{1985AJ.....90..418K,1995AJ....110.2655C, 1996ApJ...468L..77G,
1997AJ....113....1S, 1998AJ....115.1418H, 2000ApJ...536..571B, 
2000A&A...363..476B,2000AJ....119...69C,2001AJ....121.3266B, 2004PASP..116..345C}
have provided
a possible solution to the major limitation of large redshift surveys --
that they are severely limited both in depth and area by the
throughput of spectrographs. Photometric redshift algorithms essentially
define a mapping from the observed photometric properties of galaxies to
their redshifts and other physical properties such as luminosity
and type. Given an accurate photometric redshift algorithm, one can 
map the observable Universe in three dimensions just by imaging
in carefully chosen passbands. The relative efficiency of imaging compared to
spectroscopy allows one to both go deeper and cover a larger area
than is possible with traditional redshift surveys. Such surveys would 
be invaluable both for studies of large scale structure,
as well as understanding the evolution of galaxies. In addition, imaging surveys
with well understood redshift distributions are essential for
efforts to directly map the matter distribution using weak lensing.

Defining a photometric redshift catalogue involves fulfilling two requirements: one 
must photometrically specify a population of galaxies 
for which reliable photometric redshifts can
be obtained, and one must characterize the photometric redshift error
distribution. Demonstrating this process 
is the purpose of this work, using the five colour imaging of the
Sloan Digital Sky Survey \citep[SDSS,][]{2000AJ....120.1579Y} as an example.

Luminous Red Galaxies (LRGs) have long been recognized as a 
promising population for the application of photometric redshifts 
\citep{1985ApJ...297..371H,2000AJ....120.2148G, 2001AJ....122.2267E, 2001MNRAS.325.1002W}. 
These galaxies have 
remarkably uniform spectral energy distributions
\citep[SEDs,][]{1983ApJ...264..337S,2003ApJ...585..694E} that are characterized
by a strong break at 4000 \AA~caused by the accumulation of a number of metal
lines. The redshifting of this feature through different filters gives these 
galaxies their characteristic red colours that are strongly 
correlated with redshift. This makes it easy to select these galaxies
and to estimate photometric redshifts. In addition, these
are among the most luminous galaxies in the Universe, and
map large cosmological volumes. Furthermore, LRGs are 
strongly correlated with clusters, making them an ideal tool for 
detecting and studying clusters. All of the above make LRGs an astrophysically
interesting sample and an ideal candidate for a photometric redshift survey.

Measuring the photometric redshift error distribution requires a calibration set of
spectroscopic redshifts that span a similar colour and magnitude range 
as the photometric catalogue. We use two redshift catalogues to calibrate the
LRG photometric redshifts, the
SDSS Data Release 1 \citep{2003AJ....126.2081A}
\footnote{We note that Data Release 1 here only refers to the area coverage;
the reduction pipelines used are identical with those for DR2 and DR3. In 
particular, the model magnitude bug in the DR1 reductions does not affect this 
paper.}
LRG spectroscopic catalogue for redshifts $<$ 0.4, and the SDSS-2dF LRG 
spectroscopic catalogue \citep{cannon_aao_sdss2df} for redshifts between 0.4 and 0.7. These 
catalogues have extremely good coverage of the LRG colour and magnitude selection 
criteria by design; the selection criteria we use have been strongly 
influenced by both these catalogues. In addition, we 
supplement the low redshift catalogue with the SDSS MAIN galaxy catalogue 
complete to an $r$ band magnitude of 17.77.

A generic problem in interpreting analyses with
photometric redshifts is estimating the conditional probability
distribution, $P(z_{spectro} |  z_{photo})$, as this
allows us to connect the measurement -- the photometric
redshift -- with the physical quantity -- the actual redshift of the
galaxy. This ability to connect photometric redshifts with
actual redshifts is essential to theoretically interpret
results derived from photometric surveys, and generically will
be a significant source of systematic error. The simplest way to 
measure $P(z_{spectro} | z_{photo})$ is to directly measure it from a
calibration data set. Unfortunately, $P(z_{spectro} | z_{photo})$ depends
on the underlying redshift distribution, and therefore to obtain unbiased
results, the calibration data and the actual data must sample the same 
redshift distribution. This is quite often not the case, since calibration data
are drawn from heterogenous sources. We also note that simulations cannot
solve this problem, since the $P(z_{spectro} | z_{photo})$ derived will
depend on the simulated redshift distribution, which might differ 
significantly from the true distribution.

The approach that we favour in this paper is to use Bayes' theorem
to relate $P(z_{spectro} |  z_{photo})$ to 
$P(z_{photo} | z_{spectro})$, using
the true redshift distribution, $dN/dz$ of the photometric sample.
For samples selected only with an apparent magnitude cut, one can estimate
$dN/dz$ directly from the galaxy luminosity function \citep[for eg.][]{2003ApJ...595...59B}. 
This approach is significantly harder for samples, like the ones considered in this paper,
that involve multiple magnitude and colour cuts, as it involves
the joint luminosity-colour distribution functions that are poorly 
understood.

We present an alternative method to estimate $dN/dz$ in this paper, that
starts from the observation that the observed photometric redshift distribution
is just the true redshift distribution convolved with the photometric
redshift errors. Phrased as such, estimating $dN/dz$ is simply the 
problem of deconvolving the redshift errors from the measured 
redshift distribution. This problem, like all deconvolution problems, 
is ill-conditioned and must be regularized to obtain a stable solution.

This paper is organized as follows -- Sec. 2 describes the two sources
for our calibration data, the SDSS and SDSS-2dF surveys, and presents
our selection criteria for LRGs. In Sec. 3, we describe two photometric 
redshift algorithms and calibrate them against the catalogues from the 
previous section and measure the photometric redshift error 
distribution. Sec. 4 discusses using this error distribution to invert the 
observed photometric redshift distribution to reconstruct the true 
redshift distribution, while Sec. 5 summarizes our conclusions.
Whenever necessary, we have assumed a cosmology with $\Omega_{m} = 0.3$, 
$\Omega_{\Lambda}=0.7$ and $H_{0} = 100 h$ km/s/Mpc.

\section{Selecting Red Galaxies}

We start by describing the data that form our calibration
dataset, the Sloan Digital Sky Survey's spectroscopic (MAIN and Luminous
Red Galaxy) survey and the SDSS-2dF LRG survey; the reader is
referred to the appropriate technical documents 
\citep{2001AJ....122.2267E,2002AJ....124.1810S} for a more detailed
description. We then present the exact cuts used to construct our sample of 
LRGs. These are similar in spirit to those
in \cite{2001AJ....122.2267E} although they differ in detail.

Since the photometry for the two catalogues we use is from the SDSS, we 
restrict our discussion in this paper to the SDSS 5-filter photometric 
system \citep{1996AJ....111.1748F, 2002AJ....123.2121S}. 
The methods can be generalized to an arbitrary
photometric system. Except where explicitly specified, we will
use SDSS model magnitudes \citep{2002AJ....123..485S}; for instance, $g$ will 
refer to an SDSS $g$ band model magnitude. SDSS Petrosian magnitudes
will be denoted by a subscripted ``Petro'', e.g. $r_{Petro}$ is the SDSS $r$
band Petrosian magnitude. 

Finally, a comment on the magnitude system used~: it has become traditional 
to use AB magnitudes \citep{1983ApJ...266..713O} for estimating photometric
redshifts. The SDSS magnitudes are close to AB magnitudes, but differ at the 
millimag level \citep{2004AJ....128..502A}. 
The final zeropoint corrections for the SDSS have yet to be determined;
we use the best estimate of these offsets available at the time of writing.
The offsets applied are $\Delta(u,g,r,i,z) = (-0.042, 0.036, 0.015, 0.013, -0.002)$.
We note that the photometric redshifts are not very sensitive to the precise 
values of these offsets; not including them changes the measured redshifts by $\Delta z 
\sim 0.005$, completely subdominant to the photometric redshift errors.

\subsection{The SDSS Surveys}

The Sloan Digital Sky Survey (SDSS) is an ongoing survey to
image approximately $\pi$ steradians of the sky, and follow up
approximately one million of the detected objects spectroscopically. 
The imaging is carried out by drift-scanning
the sky \citep{1998AJ....116.3040G} 
in photometric conditions \citep{2001AJ....122.2129H}, 
in 5 ($ugriz$) bands using a specially 
designed wide-field camera.
Using these imaging data as a source,
objects targeted for spectroscopy \citep{2003AJ....125.2276B,2002AJ....124.1810S}
are observed with a 640 fiber
spectrograph on the same telescope. All of these data are processed
by completely automated pipelines that detect and measure photometric
properties of objects, and astrometrically calibrate the data
\citep{lupton, 2003AJ....125.1559P}.
The SDSS is close to completion, and has
had three major data releases (EDR, Stoughton et al, \citeyear{2002AJ....123..485S};
DR1, Abazajian et al,\citeyear{2003AJ....126.2081A}; 
DR2, Abazajian et al, \citeyear{2004AJ....128..502A}).
This paper will limit itself to DR1, with approximately 168,000 spectra.

The data used in this paper are from the MAIN \citep{2002AJ....124.1810S} and
Luminous Red Galaxy (LRG) \citep{2001AJ....122.2267E} surveys. The MAIN galaxy
sample is a magnitude limited survey targeting all galaxies with 
$r_{Petro} < 17.77$. The SDSS LRG sample
targets a smaller set of galaxies with $r_{Petro} < 19.5$; these
galaxies are colour selected to have strong 4000 \AA~breaks allowing a 
spectroscopic determination of their redshifts even though they are $\sim 2$
magnitudes fainter than the MAIN galaxy sample. The selection methodology
of these galaxies forms the basis both of the SDSS--2dF survey which we now 
discuss, and the selection criteria we present in Sec.\ref{sec:select}.

\subsection{The SDSS--2dF Survey}

The second set of observations are
the first data obtained as part of the SDSS-2dF LRG
survey.  This redshift survey, started in early 2003, exploits the marriage of
two facilities; the wide-angle, multi-colour, imaging data of the SDSS
and the 2dF spectrograph on the
4--meter Anglo--Australian Telescope \citep[AAT,][]{2002MNRAS.333..279L}. The SDSS--2dF LRG survey
is being carried out in tandem with  
the SDSS-2dF QSO survey to ensure optimal use of the
400 spectroscopic fibers available in the 2dF spectrograph.

The goal of the SDSS-2dF LRG survey is to replicate the selection of SDSS LRGs
but at a higher redshift, by going to fainter apparent
luminosities. In particular, we aim
to closely match the space density, luminosity range and colours of the lower
redshift SDSS LRGs, thus allowing study of the evolution of a single
population of massive galaxies over a large redshift range.
To achieve this goal, we use the same methodology as outlined in \cite{2001AJ....122.2267E}
for selecting the low redshift SDSS LRGs, but adapt the colour and
magnitude cuts to preferentially select LRGs in the redshift range $0.45 < z<
0.7$. The SDSS-2dF LRG cuts we use are similar to those of the Cut II SDSS LRG
sample discussed in detail in \cite{2001AJ....122.2267E}. However, because of
the larger telescope (AAT), and the longer integration times possible, we 
relax the $r<19.5$ magnitude limit of Cut II (which resulted in a severe
redshift limit of $z\simeq 0.45$ for the SDSS LRGs) to $i\le20$.
As discussed in \cite{2001AJ....122.2267E}, the selection of LRGs
above $z\simeq0.4$ is actually easier than selecting them at lower redshifts
because the 4000 \AA~break
 moves into the SDSS $r$ band and therefore, the SDSS $r-i$ colour
is an effective estimate of the redshift, while the $g-r$ colour is a proxy for
the rest--frame colour of the galaxy.

The details of the SDSS-2dF LRG selection criteria will be presented
elsewhere. However, as shown in \cite{2003astro.ph..5041N} and Fig. \ref{fig:catalog},
the SDSS-2dF selection
criteria successfully reproduce the luminosity range covered by the lower
redshift SDSS LRGs \citep[both Cut I and Cut II]{2001AJ....122.2267E} 
over the expected range of redshifts from $0.4<z<0.75$. 
Note that the $r-i$ colour selection is very effective at isolating 
high redshift galaxies, with 90\% of the galaxies having redshifts 
between $0.4$ and $0.7$, and virtually none with redshifts $< 0.3$.
The SDSS--2dF LRG and
QSO surveys are underway with the goal of obtaining the
final sample of $\simeq10,000$ LRGs and quasars. The data 
we use are all the data observed through 2003 with reliable spectroscopic
redshifts, a sample of $\sim 3000$ galaxies.

\subsection{Selection Criteria}
\label{sec:select}

We now discuss the construction of a photometric sample of LRGs. Although
the selection criteria we present here  (including the 
terminology) are based on the spectroscopic selection
used to construct the two samples discussed 
above, we emphasize that these are not the specific selection criteria for 
either sample, but rather are a synthesis of different
selection techniques. The goal of these selection criteria is to photometrically
select a uniform sample of LRGs over the redshift range  $0.2 < z < 0.7$.

\begin{figure}
\begin{center}
\leavevmode
\includegraphics[width=3.0in]{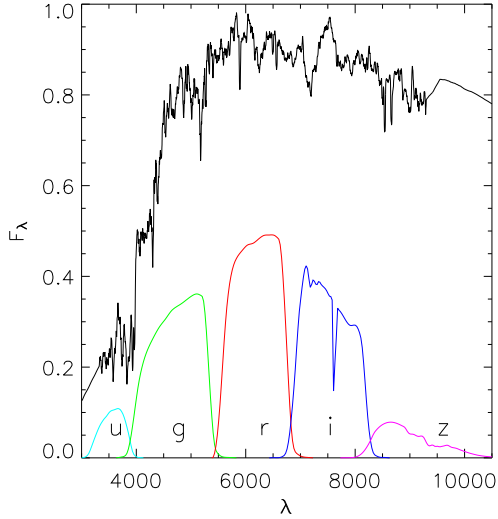}
\end{center}
\caption{A model spectrum of an early type galaxy from 
Bruzual \& Charlot (\citeyear{2003MNRAS.344.1000B}). 
The model was formed from a single burst of star formation
11 Gyr ago, and assumes a
solar metallicity. Note the prominent break in the spectrum at 4000 \AA.
Also overplotted are the response functions (including atmospheric
absorption) for the SDSS filters.}
\label{fig:modelspec}
\end{figure}

\begin{figure}
\begin{center}
\leavevmode
\includegraphics[width=3.0in]{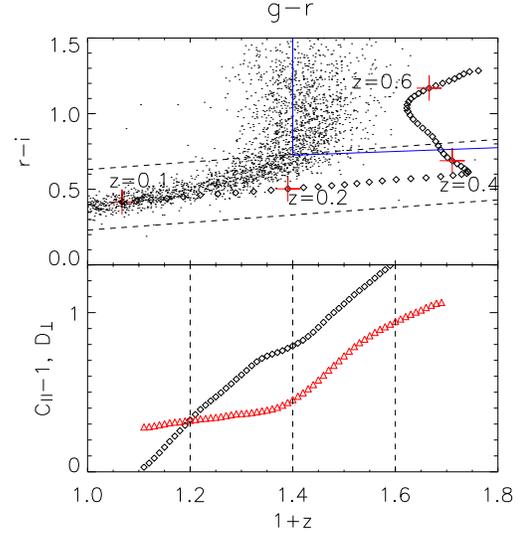}
\end{center}
\caption{The top panel shows simulated $g-r$ and $r-i$ colours of 
an early-type galaxy as
a function of redshift. The spectrum used to generate the 
track is the same as in Fig. \ref{fig:modelspec}, but evolved in 
redshift. Also shown are the colour cuts for Cut I (dashed, black)
and Cut II galaxies (solid, blue). The points show 
the stellar locus as determined by a sample of stars with $r$-band magnitudes 
less than 19.5. The lower panel shows the 
colours $c_{||}$ (diamonds, black) and $d_{\perp}$ (triangles, red),
as a function of redshift. Also shown are fiducial redshift boundaries for
Cut I (0.2 -- 0.4) and Cut II (0.4 -- 0.6). Note that the range in $g-r$
is identical to the range in $1+z$.
}
\label{fig:modelcolor}
\end{figure}

Fig. \ref{fig:modelspec} shows a model spectrum of an early type galaxy 
from the stellar population synthesis models of \cite{2003MNRAS.344.1000B}.
This particular spectrum is derived from a single burst of star formation 11 Gyr ago
(implying a redshift of formation, $z_{form} \sim 2.6$), 
evolved to the present, and is typical of LRG spectra. In particular, the
4000 \AA~break is very prominent.
In order to motivate our selection criteria, we passively
evolve this spectrum in redshift (in particular, 
taking the evolution of the strength of the 4000 \AA break
into account), and project it through the SDSS filters; the 
resulting colour track in $g-r-i$ space as a function of redshift is shown in
Fig. \ref{fig:modelcolor}. The bend in the track around $z \sim 0.4$, caused 
by the redshifting of the 4000 \AA~break from the $g$ to $r$ band, naturally 
suggests two selection criteria -- a low redshift sample (Cut I), nominally from
$z \sim 0.2 - 0.4$, and a high redshift sample (Cut II), from $z \sim 
0.4 - 0.6$. We define the two colours 
\citep[][and private commun.]{2001AJ....122.2267E}
\bea
c_{\perp} \equiv (r-i) - (g-r)/4 - 0.18 \,\,\, ,\\
d_{\perp} \equiv (r-i) - (g-r)/8 \approx r-i \,\,\,.
\label{eq:perpdef}
\eea
We now make the following colour selections,
\bea
{\rm Cut\,\,I :} & \mid c_{\perp} \mid < 0.2 \,\,\,;\\
{\rm Cut\,\,II :} & d_{\perp} > 0.55 \,\,\,, \\
& g-r > 1.4 \,\,\,,
\label{eq:colourcuts}
\eea
as shown in Fig. \ref{fig:modelcolor}. The final cut, $g-r > 1.4$, 
isolates our sample from the stellar locus. In addition to these selection
criteria, we eliminate all galaxies 
with $g-r > 3$ and $r-i > 1.5$; these constraints
eliminate no real galaxies, but are effective in removing stars with 
unusual colours.

Unfortunately, as emphasized in \cite{2001AJ....122.2267E}, these simple colour
cuts are not sufficient to select LRGs due to an accidental degeneracy 
in the SDSS filters that causes all galaxies, irrespective 
of type, to lie very close to the low redshift early type locus. We therefore
follow the discussion there and impose a cut in absolute magnitude. We implement
this by defining a colour to use as a proxy for redshift and then translating
the absolute magnitude cut into a colour-magnitude cut. We see from Fig. 
\ref{fig:modelcolor} that $d_{\perp}$ correlates strongly with redshift and
is appropriate to use for Cut II. For Cut I, we define,
\be
c_{||} = 0.7 (g-r) + 1.2(r-i-0.18) \,\,\,,
\label{fig:cpllel}
\ee
which is approximately parallel to the low redshift locus. Given these, we 
further impose
\bea
{\rm Cut\,\,I :} & r_{Petro} < 13.6 + c_{||}/0.3 \,\,\,,\nonumber \\
& r_{Petro} < 19.7 \,\,\,; 
\label{eq:colourmagcuts1}
\eea
\bea
{\rm Cut\,\,II :} & i < 18.3 + 2d_{\perp} \,\,\,, \nonumber \\
& i < 20 \,\,\,.
\label{eq:colourmagcuts}
\eea
Note we use $r_{Petro}$ for consistency with the SDSS LRG
target selection.
We note that Cut~I is identical (except in the numerical values
of the magnitude cuts in Eqs. \ref{eq:colourmagcuts1}) to the 
SDSS LRG Cut~I, while the numerical values for Cut~II were chosen to yield a 
population consistent with Cut~I (see below). 
This was intentionally done to maximize the overlap between 
any sample selected using these cuts, and the SDSS LRG spectroscopic sample.
The switch to the $i$ band for Cut~II also requires some explanation. As is
clear from Fig.\ref{fig:modelspec}, the 4000 \AA~break is moving through 
the $r$ band throughout the fiducial redshift range of Cut II. This implies 
that the K-corrections to the $r$ band are very sensitive to redshift; the $i$ band
K-corrections are much less sensitive to redshift allowing for a more
robust selection.

\begin{figure}
\begin{center}
\leavevmode
\includegraphics[width=3.0in]{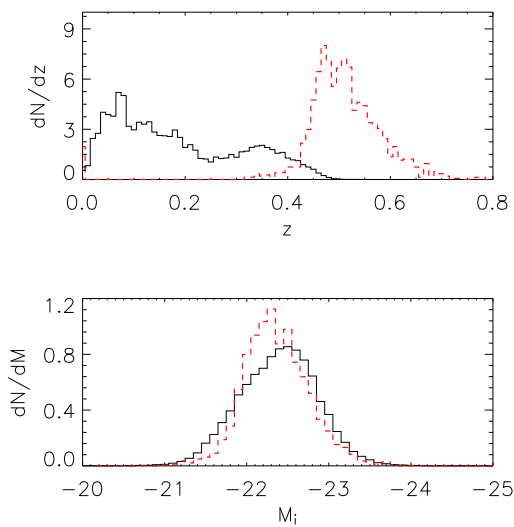}
\end{center}
\caption{The top panel shows the spectroscopic redshift distribution, 
$dN/dz$ of the SDSS (solid, black) and the SDSS-2dF (dashed, red) samples
trimmed using the selection criteria of Sec.\ref{sec:select}. Note that the
SDSS sample is dominated by the low redshift MAIN sample, accounting for 
the low normalization at high redshift. The lower panel shows the 
$i$ band absolute magnitude distribution for the two samples, demonstrating that
our absolute magnitude cuts are selecting a sample with $M_{i} \sim -22$ 
as desired. Both $dN/dz$ and $dN/dM$ are normalized so that they integrate
to unity.
}
\label{fig:catalog}
\end{figure}

The results of applying these cuts to the spectroscopic catalogs
are shown in Fig. \ref{fig:catalog}. Since
the SDSS spectroscopic catalogue is at low redshift, we trim the
catalogue using Cut~I, 
while the higher redshift SDSS-2dF data are trimmed with Cut~II. 
Our calibration dataset has 45,744 Cut~I galaxies, and 1,474 Cut~II
galaxies. The large number of low redshift galaxies that pass Cut~I indicate
a failure of our selection criteria at redshifts lower than $z \sim 0.15$, 
as already noted by \cite{2001AJ....122.2267E}. We however leave these
galaxies in our analysis, since they will contaminate any photometrically selected
sample and it will be necessary to understand their photometric redshift 
distributions. No such problem exists for the Cut II galaxies, which
have a negligible fraction of $z < 0.4$ galaxies. The most significant contaminant
for Cut~II are M stars. The $g-r > 1.4$ cut
removes most of these, although there is a small residual level 
of contamination. Analyses using this or similar samples will have to
estimate the effect of this contamination on their results.

The lower panel of Fig. \ref{fig:catalog} shows the absolute magnitude
distribution of both Cut I and Cut II galaxies. As expected, the 
colour magnitude cuts restrict the sample to bright galaxies; the 
median $i$ band magnitude is $M_{i} \sim -22$. Note that the Cut~I and
Cut~II galaxies probe similar luminosities.
The Cut~I magnitude distribution also
has a tail extending to low luminosities; this is the failure of the 
selection criteria at low redshift we encountered above.

\section{Photometric Redshift Estimation}

Photometric redshift estimation techniques can be classified into
two groups, ``empirical'' and ``template-fitting'' methods. Empirical methods
\citep{1995AJ....110.2655C,1999ApJ...516..563B,1998AJ....116.2081W}
are based on the observational fact that galaxies are 
restricted to a low dimensional surface in the space of their colours
and redshift. Using a training set of galaxies, these methods attempt to
parametrize this surface, either with low-order polynomials, nearest-neighbour
searches or neural networks. The advantage of these methods is that they
attempt to measure these relationships directly from the data, and so, 
implicitly correct for any calibration biases present. A publicly available 
example is the Artificial Neural Network 
code \citep[ANNz,]{2003MNRAS.339.1195F,2004PASP..116..345C}
that trains a neural network to learn the relation 
between photometry and redshift from an appropriate training set
of observed galaxies whose redshifts are already known. This code
has a photometric redshift accuracy similar to the methods 
decribed below\citep{2004PASP..116..345C}.

The fact that these methods rely on training sets is their
greatest disadvantage. For these methods to work, the training set must 
densely sample the entire redshift-colour space of interest, as it is
difficult to extrapolate outside the domain of the training set. Most
training sets, including the samples constructed above, violate the 
above requirement and therefore are of limited utility. 
Template-based methods do not suffer from these drawbacks, and form the basis for 
the two algorithms used in this paper, which we now discuss.

\subsection{Simple Template Fitting}
\label{sec:single}

Template fitting methods start with a set of model spectra (the ``templates'')
of galaxies, either from spectrophotometrically calibrated observations of
galaxies \citep{1980ApJS...43..393C} or from stellar synthesis models 
\citep{2003MNRAS.344.1000B,2002A&A...386..446L}.
These methods then attempt to reconstruct the observed 
colours of galaxies by some (appropriately redshifted) linear combination of the
templates, projected through the appropriate filters. The best fit redshift
is then an estimate of the galaxy's true redshift.
Concretely, if we denote the templates by $\Psi^{i}(z)$, this algorithm
can be cast as a minimization of 
\be
\chi^{2}(c_{i},z) = \sum_{\alpha} \left( \frac{f_{\alpha} - 
R_{\alpha}(\sum_{i} c_{i} \Psi^{i}(z))}{\sigma_{\alpha}} 
\right)^{2} \,\,\, ,
\ee
where $f_{\alpha}$ is the observed flux (with error
$\sigma_{\alpha}$) of the galaxy in the $\alpha$ filter, and 
$R_{\alpha}(\Psi)$ projects the spectrum $\Psi$ onto the $\alpha$ filter.
For definiteness, we work with the AB photometric system \citep{1983ApJ...266..713O}, 
where the apparent magnitude of a galaxy, $m_{AB}$, is related to its
SED, $\Psi$ (with units of W m$^{-2}$ Hz$^{-1}$), by
\be
m_{AB} = -2.5 \log_{10} \left[\frac{\int \, d\nu \, \nu^{-1} \Psi(\nu) W_{\alpha}(\nu)}
{\int \, d\nu \, \nu^{-1} g(\nu) W_{\alpha}(\nu)}\right] \,\,
\ee
where $W_{\alpha}$ is the response of the $\alpha$ filter.
The reference SED is given by $g(\nu) = 3631 \rm{Jy}$ (where
1 Jy = $10^{-26}$ W m$^{-2}$ Hz$^{-1}$).

One of the advantages of the LRGs is that their spectra are well
described by a single template \citep{2003ApJ...585..694E}.
We find that the LRG colours are well described by a \cite{2003MNRAS.344.1000B}
\footnote{These template stellar population spectra are part of 
the GALAXEV package available at \texttt{http://www.cida.ve/$\sim$bruzual/bc2003}.}
single instantaneous burst template at solar metallicity with the burst
occurring when the Universe was 2.5 Gyr old. The template evolves with time, 
becoming redder and increasing the 4000 \AA~break, as the more massive hot stars
die.
To incorporate this evolution, we interpolate between
models with bursts with ages of 
[11, 5, 2.5, 1.4, 0.9, 0.64, 0.1] Gyr  
to calculate the template as a function
of redshift. The photometric redshifts we derive are insensitive to 
the precise time of the burst, and therefore, we do not attempt any optimization of
this parameter.

The implementation of this method we use is part of the IDLSPEC2D SDSS 
spectroscopic reduction pipeline \citep{schlegel} and can be downloaded 
through the WWW\footnote{\texttt{http://spectro.princeton.edu}}.

\subsection{Hybrid Methods}
\label{sec:hybrid}

Obviously, this simple template-fitting algorithm is effective only when
the templates accurately describe the photometric properties of the
galaxies for which one wants to estimate redshifts. This suggests generalizing 
the template-fitting algorithm by incorporating features of empirical
methods \citep{2000AJ....119...69C,2000AJ....120.1588B,2001AJ....121.3266B}.
The basic approach
is to divide a training set into spectral classes corresponding to 
each of the templates. Given these training sets for the individual templates,
one can {\it repair} the templates by adjusting them to better reproduce the
measured colours of the galaxies in the training set. By repeating this 
classification and repair procedure, one can obtain a improved template set 
that yields more reliable photometric redshifts \citep{2003AJ....125..580C}.
Moreover, this process of adjusting the templates to agree with observations
makes hybrid methods potentially less sensitive to potential systematic problems due
to errors in the filter curves or photometric zeropoints. 
We refer the reader to the above papers for details on the implementation of this
algorithm. For the LRGs, we start with an elliptical template from 
\cite{1980ApJS...43..393C} and apply the above algorithm to 
optimize it. This is done in an iterative manner and converges after 
typically three iterations.

This single optimized template is used for an initial            
redshift estimate for all LRGs. The SDSS LRG sample is, however,        
selected assuming a passively 
evolving elliptical template \citep{2001AJ....122.2267E}. 
Therefore, we expect to gain in photometric             
redshift accuracy if we allow the LRG spectral template to evolve with          
redshift. The SDSS and SDSS-2dF redshift samples are subdivided          
into three redshift intervals $0.00<z<0.35$, $0.25<z<0.45$ and                  
$0.35<z<1.0$, based on the photometric redshifts of the individual              
galaxies. Within each interval we optimize the spectral template as             
described above. The overlapping redshift intervals provide a smooth            
progression in spectral type from one redshift interval to the next, as          
well as ensuring that the number and distribution of calibration                
redshifts is sufficient to constrain the colours of galaxies across a            
broad spectral range. 

\subsection{Results}
\label{sec:results}

\begin{figure}
\begin{center}
\leavevmode
\includegraphics[width=3.0in]{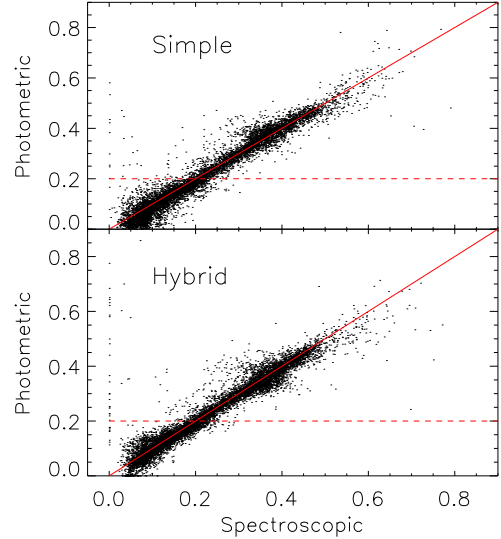}
\end{center}
\caption{Scatter plot showing the photometric redshift versus
the spectroscopic redshift for a random 10000 galaxies from 
our calibration sample. The upper panel shows the results
for the simple template fitting code of Sec.\ref{sec:single}, and
the lower panel are the results for the hybrid code of
Sec.\ref{sec:hybrid}. The solid (red) line has slope 1, while
the dashed line marks the fiducial lower redshift limit of any
photometric LRG sample. 
The difficulty of estimating redshifts at $z\sim 0.4$ is evident from
the increased scatter.
}
\label{fig:photoz_scatter}
\end{figure}

We now apply the methods of the previous two sections to our calibration
dataset; the results are summarized in Fig.\ref{fig:photoz_scatter}. Both
are essentially unbiased ($|\Delta z_{mean}| < 0.01$) at redshifts less that 0.5. At 
higher redshifts, the photometric redshifts are systematically lower than 
the spectroscopic redshifts by about 5\%. The scatter in both methods is 
approximately $\sigma \sim 0.035$, except at redshifts greater than $0.55$, 
where the scatter grows to $\sim 0.06$, caused both by increased photometric
scatter and increased uncertainties in templates (caused by for eg. star formation
or emission lines).

There are two noticeable features in Fig.\ref{fig:photoz_scatter} that deserve
comment. The first is that the hybrid methods do significantly better than the
single template fits at low redshifts ($z < 0.15$). This is due to 
the failure of the LRG selection criteria at low redshifts; a single 
elliptical template no longer well describes this population. This highlights 
an important advantage of the hybrid methods -- they adjust their
templates to better describe the populations.

\begin{figure}
\begin{center}
\leavevmode
\includegraphics[width=3.0in]{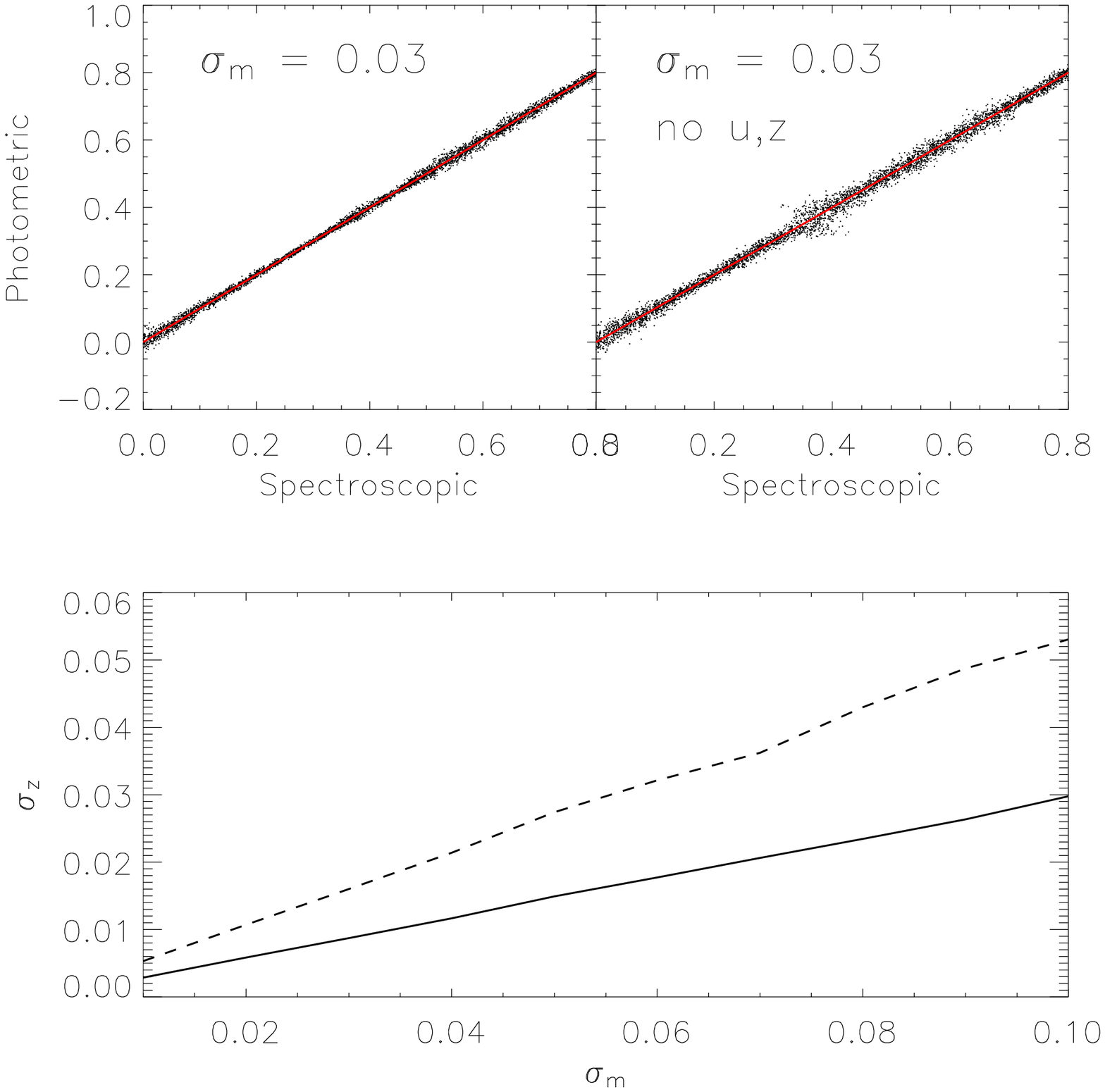}
\end{center}
\caption{Simulations showing the effect of magnitude errors
on the accuracy of the photometric redshifts. The upper left
plot shows the reconstructed photometric redshifts for a 
magnitude error, $\sigma_{m} = 0.03$ in all 5 bands, while the 
upper right panel has no S/N in the $u$ and $z$ bands and $\sigma_{m} = 0.03$
in the remaining bands. The lower panel shows the redshift error
induced by magnitude errors; the solid line has constant error across 
the bands, while the dashed line has constant error in $g,r,i$ and 
zero S/N in $u$ and $z$. Since the magnitude errors are independant of 
redshift, the redshift errors are simply computed over the entire 
redshift range.
}
\label{fig:magerrors}
\end{figure}

The second feature is the increased scatter around $z \sim 0.4$,
caused by an accidental degeneracy due to 
the SDSS filters. Fig. \ref{fig:modelspec}
shows a gap between the $g$ and $r$ bands 
at about 5500 \AA\footnote{This gap is partly intentional, 
avoiding the OI (5577~\AA) 
night-sky emission lines. However, the filters
were intended to overlap more.}.
As the 4000 \AA~break enters this gap at $z \sim 0.38$, the lack of 
coverage in either the $g$ or $r$ band causes a degeneracy between the
strength of the 4000 \AA~break and its location, increasing the
redshift errors.

\begin{figure*}
\begin{center}
\leavevmode
\includegraphics[width=6.0in]{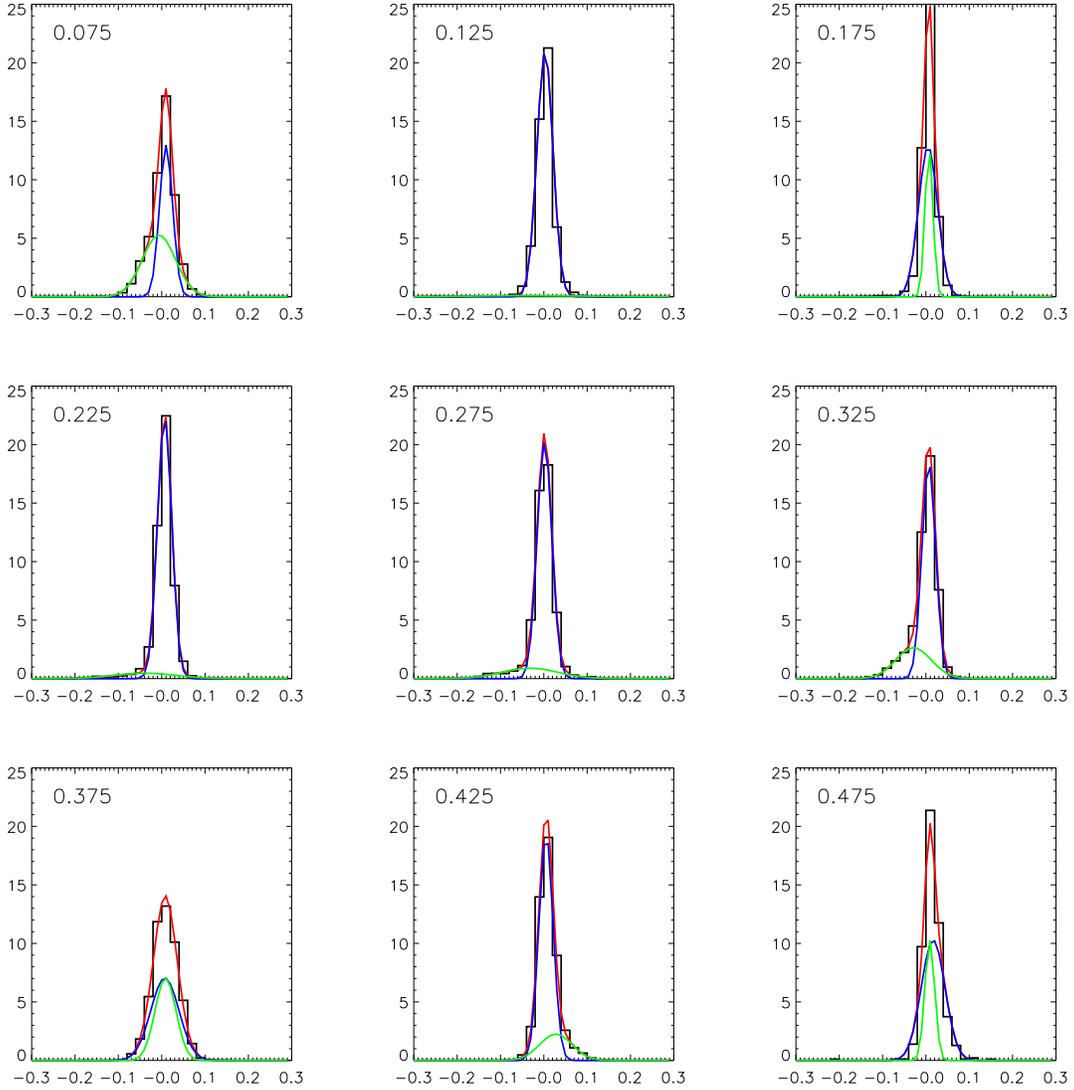}
\end{center}
\caption{The double Gaussian fits to the error distribution as a function
of spectroscopic redshift. The $x$ axis shows the redshift error, $z_{spectro} -
z_{photo}$, and each panel is a redshift slice with the central redshift
shown in the upper left. The histogram is the measured distribution, while the 
curves are the best fit Gaussian (both individually and summed). The data here
are SDSS galaxies selected using Cut I. The photometric redshifts are estimated
using the method of Sec. \ref{sec:hybrid}.
}
\label{fig:sdss_2gauss}
\end{figure*}

It is useful to be able to separate the effects of template errors from 
photometric errors in the redshift error budget. In order to 
do this, we simulate galaxies by uniformly distributing them between 
$0 < z < 0.8$ with
synthetic colours given by the template used in the method of 
Sec.\ref{sec:single}. We then add errors to these synthetic fluxes; focussing
on two extreme cases -- uniform errors across all 5 bands, and no S/N in 
the $u$ and $z$ band (i.e. infinite magnitude errors corresponding 
to non-detection in $u$ and $z$ band, with uniform errors in the other bands). The latter
case is motivated by the fact that the SDSS camera is least sensitive in 
the $u$ and $z$ bands, and because most LRGs are not detected 
in the $u$ band. The results of this exercise are shown in Fig. \ref{fig:magerrors}.
The upper panels show a realization with a (optimistic) magnitude error, $\sigma_m \sim 0.03$.
For comparison, the median S/N ($\sim 1/\sigma_{m}$)
for LRGs at $z\sim 0.3$ is $\sim (2,30, 70, 80, 30)$, and
$\sim (0.5, 10, 25, 36, 15)$ at $z \sim 0.5$.
A prominent feature is the degeneracy at $z\sim 0.4$ discussed above,
for the case where the $u$ and $z$ bands have
no signal. 
In this case, there is no extra information that can be used
to break the degeneracy between the 4000 \AA~break strength and its location.
Also, the scatter in the photometric redshifts increases for redshifts 
greater than $\sim 0.35$, coinciding with the 4000 \AA~break moving into the $r$
band, and the loss of redshift information from the $g-r$ colour.
The lower panel shows how the redshift errors increase with increasing magnitude
errors, again for the cases of uniform S/N in all bands, and in $g,r,i$ with 
zero S/N in $u$ and $z$. 
We also note that the redshift errors we measure are consistent with being caused
principally by photometric scatter. However, the bias seen at high redshifts cannot
be caused by photometric errors, and suggest either errors in the template, or 
errors in the photometric zeropoints or filter curves.

\begin{figure*}
\begin{center}
\leavevmode
\includegraphics[width=6.0in]{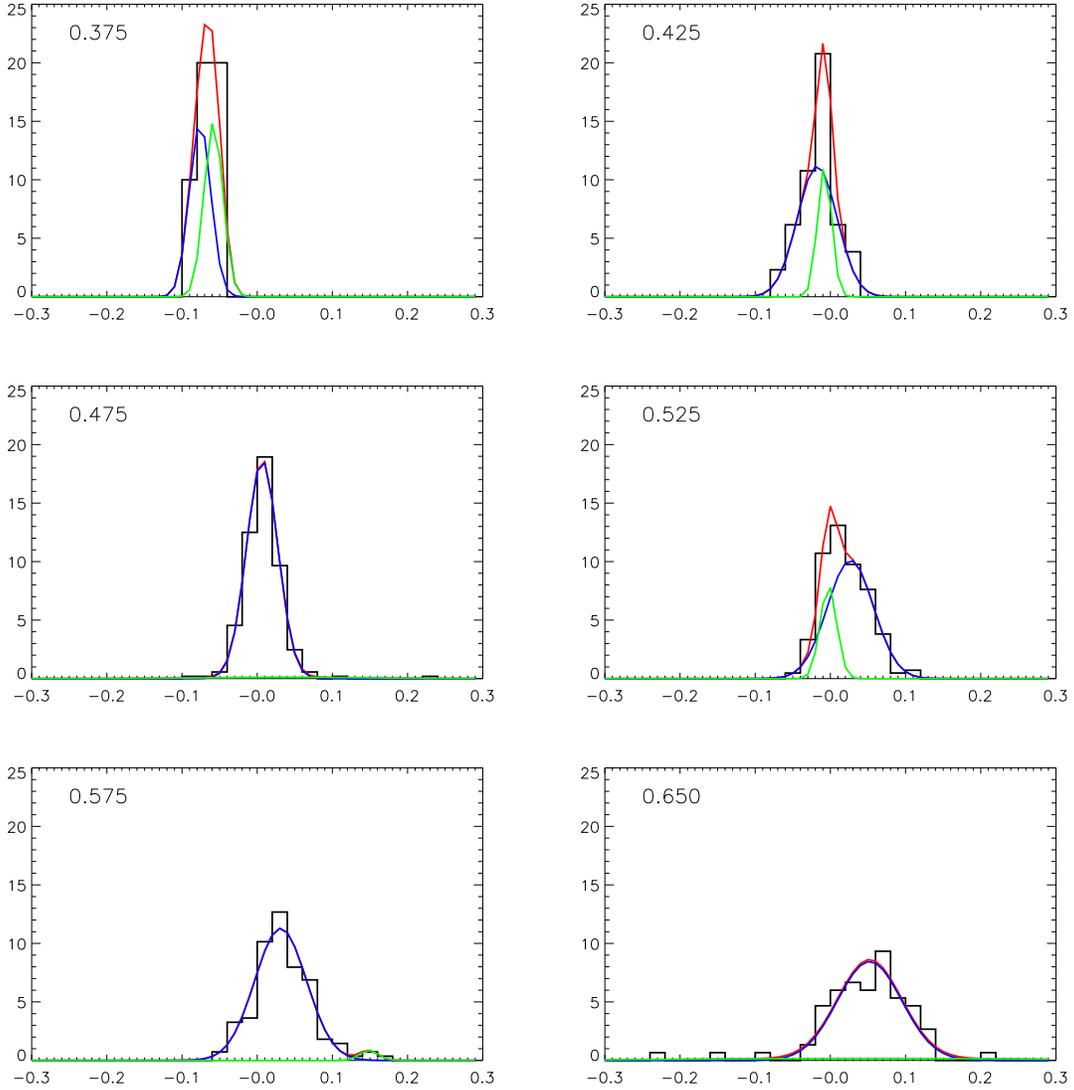}
\end{center}
\caption{Same as Fig. \ref{fig:sdss_2gauss} except for the SDSS-2dF galaxies
selected using Cut II and from redshifts 0.35 to 0.7.
}
\label{fig:2df_2gauss}
\end{figure*}

In order to parametrize the error distribution, we divide the 
calibration data into redshifts of width 0.05 (except between $z=0.6$ 
and $0.7$, which we combine because of the small number of galaxies
in that range). Within each of these ranges, we fit the error distribution
with a sum of Gaussians,
\be
P(\delta z = z_{\rm spectro} - z_{photo}) = \frac{1}{{\cal N}} \sum_{i=1}^{N} 
  b_{i} \exp\left(\frac{-(\delta z - m_{i})^{2}}{2 \sigma_{i}^2}\right) \,\,\,,
\label{eq:errormodel}
\ee
where ${\cal N}$ is the normalization given by
\be
{\cal N} = \sum_{i=1}^{N} \sqrt{2 \pi} b_{i} \sigma_{i} \,\,\,.
\ee
We find, as shown in Figs. \ref{fig:sdss_2gauss} and \ref{fig:2df_2gauss}, that 
the error distribution is well approximated by two Gaussians. The parameters of 
the fits for both the simple template and hybrid methods
are in Tables \ref{tab:error_2gauss} and \ref{tab:error_2gauss_connolly} respectively.
We note that the cores of these error distributions are significantly tighter 
than the errors mentioned above. However, the error distributions
typically have long wings that
are responsible for most of the measured RMS errors. The discrepancies between the
SDSS and SDSS-2dF samples in the overlap region are due to a colour bias introduced
by the sharp colour cuts, resulting in a bias in the redshift estimation for Cut II
galaxies between $z=0.35-0.45$. We therefore recommend using the SDSS distributions
to $z=0.45$ for samples by combining Cut~I and Cut~II.

\begin{table}
\begin{tabular}{lcccccc} 
\multicolumn{7}{c}
{SDSS/SDSS-2dF Photometric redshift errors} \\ 
\multicolumn{7}{c}
{Single Template Fitting} \\
\multicolumn{7}{c}
{Double Gaussian fits} \\
Catalogue & $z_{bin}$ & $m_{1}$ & $\sigma_{1}$ & $m_{2}$ & $\sigma_{2}$ & $b$ \\ 
\hline
SDSS &0.05-0.10 &    0.031 &    0.045 &   -0.194 &    0.084 &    0.002 \\ 
SDSS &0.10-0.15 &    0.008 &    0.019 &    0.029 &    0.060 &    0.119 \\ 
SDSS &0.15-0.20 &    0.010 &    0.015 &    0.013 &    0.052 &    0.059 \\ 
SDSS &0.20-0.25 &    0.008 &    0.036 &    0.011 &    0.013 &    1.000 \\ 
SDSS &0.25-0.30 &    0.002 &    0.018 &   -0.026 &    0.070 &    0.058 \\ 
SDSS &0.30-0.35 &   -0.002 &    0.019 &   -0.030 &    0.040 &    0.338 \\ 
SDSS &0.35-0.40 &   -0.004 &    0.028 &    0.000 &    0.000 &    1.000 \\ 
SDSS &0.40-0.45 &    0.007 &    0.028 &   -0.003 &    0.014 &    1.000 \\ 
SDSS &0.45-0.50 &    0.015 &    0.023 &    0.001 &    0.011 &    1.000 \\ 
SDSS-2dF &0.35-0.40 &   -0.064 &    0.026 &   -0.069 &    0.002 &    1.000 \\ 
SDSS-2dF &0.40-0.45 &   -0.019 &    0.022 &   -0.330 &    0.000 &    0.000 \\ 
SDSS-2dF &0.45-0.50 &    0.003 &    0.030 &    0.010 &    0.016 &    1.000 \\ 
SDSS-2dF &0.50-0.55 &    0.018 &    0.035 &   -0.278 &    0.000 &    0.000 \\ 
SDSS-2dF &0.55-0.60 &    0.023 &    0.049 &    0.066 &    0.113 &    0.021 \\ 
SDSS-2dF &0.60-0.70 &    0.039 &    0.047 &    0.011 &    0.093 &    0.423 \\ 

\hline
\end{tabular}
\caption{Double Gaussian fits to the 
photometric redshift error, $z - z_{p}$ as a
function of $z$ for the SDSS and SDSS-2dF data. $(m_{1}, \sigma_{1})$ and 
$(m_{2}, \sigma_{2})$ are the mean and standard deviation of the first and 
second Gaussians respectively, while $b$ is the ratio of the amplitude of the
second Gaussian to the first. The photometric redshifts were computed using the
method of Sec.\ref{sec:single}. We recommend using the SDSS distributions to 
$z=0.45$ and SDSS-2dF for higher redshifts.}
\label{tab:error_2gauss}
\end{table}

\begin{table}
\begin{tabular}{lcccccc}
\multicolumn{7}{c}
{SDSS/SDSS-2dF Photometric redshift errors} \\ 
\multicolumn{7}{c}
{Hybrid method} \\
\multicolumn{7}{c}
{Double Gaussian fits} \\
Catalogue & $z_{bin}$ & $m_{1}$ & $\sigma_{1}$ & $m_{2}$ & $\sigma_{2}$ & $b$ \\
\hline
SDSS &0.05-0.10 &    0.011 &    0.014 &   -0.005 &    0.036 &    0.491 \\ 
SDSS &0.10-0.15 &    0.001 &    0.027 &    0.005 &    0.011 &    1.000 \\ 
SDSS &0.15-0.20 &    0.005 &    0.022 &    0.007 &    0.009 &    1.000 \\ 
SDSS &0.20-0.25 &    0.008 &    0.016 &   -0.040 &    0.060 &    0.025 \\ 
SDSS &0.25-0.30 &    0.002 &    0.017 &   -0.028 &    0.063 &    0.050 \\ 
SDSS &0.30-0.35 &    0.006 &    0.016 &   -0.029 &    0.041 &    0.136 \\ 
SDSS &0.35-0.40 &    0.006 &    0.032 &    0.011 &    0.026 &    1.000 \\ 
SDSS &0.40-0.45 &    0.014 &    0.027 &    0.004 &    0.012 &    1.000 \\ 
SDSS &0.45-0.50 &    0.019 &    0.025 &    0.008 &    0.011 &    1.000 \\ 
SDSS-2dF &0.35-0.40 &   -0.076 &    0.014 &   -0.058 &    0.012 &    1.000 \\ 
SDSS-2dF &0.40-0.45 &   -0.017 &    0.026 &   -0.008 &    0.009 &    1.000 \\ 
SDSS-2dF &0.45-0.50 &    0.007 &    0.021 &    0.037 &    0.098 &    0.006 \\ 
SDSS-2dF &0.50-0.55 &    0.027 &    0.031 &   -0.003 &    0.011 &    0.792 \\ 
SDSS-2dF &0.55-0.60 &    0.031 &    0.034 &    0.148 &    0.013 &    0.075 \\ 
SDSS-2dF &0.60-0.70 &    0.052 &    0.043 &    0.053 &    0.227 &    0.020 \\ 

\hline
\end{tabular}
\caption{Same as Table \ref{tab:error_2gauss} except that the photometric redshifts 
were computed using the methods of Sec.\ref{sec:hybrid}}
\label{tab:error_2gauss_connolly}
\end{table}

In addition to measuring the error distribution, it is useful to measure the fraction of 
galaxies whose redshifts are ``catastrophically'' wrong. We define a catastrophic failure
as a photometric redshift that differs from the spectroscopic redshift by more than 
$\Delta z_{c}$, where we use $\Delta z_{c} = 0.1$ and $0.2$. For $\Delta z_{c} = 0.1$, 
we have a catastrophic failure rate of 3.5\% for the simple template fitting algorithm,
and 1.5\% for the hybrid algorithm. However, a large fraction of this is dominated by
the underestimation of the photometric redshifts at $z > 0.5$. If we increase $\Delta z_{c}$
to 0.2, the failure rate drops to under 0.5\%.

\section{Estimating \dndz}

In the previous section, we estimated the \photoz error distributions as a function of
the true redshift of the galaxy. With this in hand, we turn to the problem of estimating
the actual redshift distribution, \dndz of a sample of galaxies given the 
distribution of their photometric redshifts, \dndzp. 

As discussed in the Introduction, 
the apparently trivial solution to this problem is to measure the error distribution
not as a function of the true redshift, but as a function of photometric redshift.
One can then add these distributions, weighted by \dndzp to estimate the true
redshift distribution.
The problem with this approach is that the \photoz error distributions measured as a function
of photometric redshift depend sensitively on the selection criteria of the 
calibration sample. If these criteria don't match those of the full sample (and in 
general, they will not), then \dndz estimated using the above technique will be biased.

In order to proceed, we observe that the photometric redshift distribution is simply
the convolution of the true redshift distribution with redshift errors,
\be
\left[\frac{dN}{dz}\right]_{P} \sim \left(\frac{dN}{dz}\right)
\otimes \Delta z \,\,.
\label{eq:dndz_schema}
\ee
If we define $\Delta(z - z_{p}, z)$ as the probability that a galaxy at redshift $z$
is scattered to photometric redshift $z_{p}$, then we can write out the above more
concretely, 
\be
\left[\frac{dN}{dz}\right]_{P}(z_{p}) = \int^{\infty}_{0} dz'\, \left[\frac{dN}{dz}\right](z') 
\Delta(z'-z_{p},z') \,\,,
\label{eq:dndz_fred}
\ee
where the left side has the known \dndzp, while the right is the unknown \dndz.
Eq. \ref{eq:dndz_fred} is a Fredholm equation of the first kind 
\footnote{For a non-technical introduction, 
see \cite{1992nrfa.book.....P}, Chap. 18} 
and is ubiquitous throughout astronomy \citep{1986ipag.book.....C}. 
Unfortunately, such problems do not possess a unique solution and moreover, are ill-conditioned.
Small perturbations in the data can produce solutions 
that are arbitrarily different. This is not surprising, given that Eq.\ref{eq:dndz_fred} 
describes a smoothing operator that generically loses information, implying that the solution
will in general require incorporating some ``prior'' knowledge about \dndz.

\subsection{Discretization and The Classical Solution}

We begin by approximating \dndzp as a stepwise constant function measured in $n$ bins, $[z_{p}^{i},
z_{p}^{i+1})$ with $i = 0, \ldots n-1$, and \dndz in $m$ bins, $[z^{j}, z^{j+1})$ 
where $j=0,\ldots, m-1$.
Substituting into Eq.\ref{eq:dndz_fred}, we obtain
\be
\left[\frac{dN}{dz}\right]_{P,i} = A_{ij} \left[\frac{dN}{dz}\right]_{j}
\label{eq:dndz_matrix}
\ee
where we assume the Einstein summation convention. The response matrix $A_{ij}$ is given by 
\be
A_{ij} = \frac{1}{z_{p}^{i+1} - z_{p}^{i}} \int^{z_{p}^{i+1}}_{z_{p}^{i}} \, dz_{p}' 
\int^{z^{j+1}}_{z^{j}} \, dz' \Delta(z'-z_{p}',z') \,\, .
\label{eq:response_matrix}
\ee
For the specific case where $\Delta$ can be described by a sum of $N$ Gaussians 
(Eq. \ref{eq:errormodel}), one can do one
of the integrals explicitly to obtain
\bea
A_{ij} = \frac{1}{z_{p}^{i+1} - z_{p}^{i}} \int^{z_{p}^{i+1}}_{z_{p}^{i}} \, dz_{p}' \,
\sqrt{\frac{\pi}{2}} \sum_{k=1}^{N} b_{k} \sigma_{k} \times \nonumber \\ 
\left[ 
f(\bar{z}_{k}, z^{j+1}, \sigma_{k}, z_{p}') - f(\bar{z}_{k}, z^{j}, \sigma_{k}, z_{p}') \right] \,\,,
\eea
where we define
\be
f(\bar{z}, z^{j}, \sigma,z) = \erf\left(\frac{|z^{j} - \bar{z} - z|}{\sqrt{2} \sigma}\right) 
\sgn \left(\frac{z^{j} - \bar{z} - z}{\sqrt{2} \sigma}\right) \,\,\,.
\ee
where $\sgn$ is the sign operator and $\erf$ is the error function.
Note that discretizing the problem has recast an integral equation (Eq.\ref{eq:dndz_fred}) 
into a matrix problem (Eq. \ref{eq:dndz_matrix}), albeit with a non-square matrix. 
We can obtain a solution to this problem by singular value decomposition
\citep[SVD,][]{1992nrfa.book.....P}.
We denote this the {\it classical} solution since we do not explicitly use any 
prior information about \dndz.

\begin{figure}
\begin{center}
\leavevmode
\includegraphics[width=3.0in]{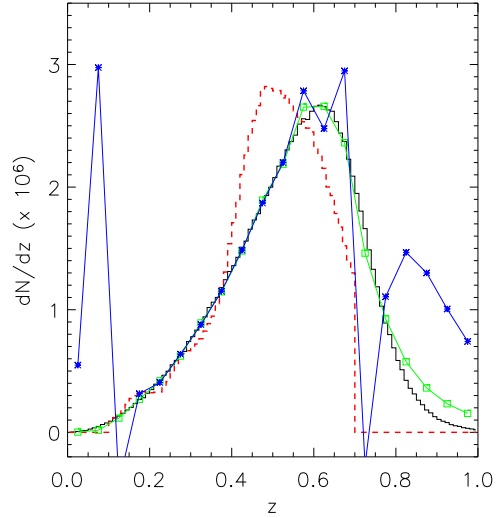}
\end{center}
\caption{Results of simulations of the classical solution of the 
redshift inversion problem. The solid [black] histogram is the true 
redshift distribution, while the broken [red] histogram shows the 
photometric redshift distribution. The connected boxes [green] and 
stars [blue] show the reconstructed redshift distributions for different
discretizations (10 and 15 bins, respectively) of the photometric
redshift distributions. In both cases, the reconstructed distribution
is parametrized by 20 step functions.
}
\label{fig:svd_inversion}
\end{figure}

In order to understand the behaviour of the classical solution, we test it 
on simulations of the photometric redshift distribution. We start by
distributing galaxies randomly in redshift between $z=0$ and $z=1$ according to, 
\be
\frac{dN}{dz} = \frac{z^{2}}{1 + \exp(20z - 14)} \,\,.
\label{eq:simulated_zdist}
\ee
This distribution initially grows as $z^2$, and is exponentially cut off 
at $z\sim 0.6$, and approximates a volume limited distribution with 
a magnitude limit at high redshifts.
Random redshift errors, using the model of Table \ref{tab:error_2gauss},
are added to obtain photometric redshifts, $z_{p}$. For redshifts greater
than 0.7, the errors are sampled from a Gaussian whose mean and width
are obtained by linearly extrapolating the errors from Table \ref{tab:error_2gauss}.
Finally, we restrict
to galaxies with $z_{p} \in [0.1, 0.7]$. An example of the true and photometric
redshift distributions is shown in Fig. \ref{fig:svd_inversion}.

The photometric redshift distribution is then discretized into $n$ bins, $[dN/dz]_{P,i}$.
We present results for $n=10\, (\Delta z=0.06)$ and $n=15\, (\Delta z=0.04)$. The estimated
$dN/dz$ is likewise parametrized as a piecewise constant function from $z=0$ to $z=1$ with
a step width of $\Delta z=0.05$. Using these parametrizations, we construct the response
matrix $A_{ij}$ (Eq.\ref{eq:response_matrix}) and solve for $dN/dz$ using Eq. \ref{eq:dndz_matrix}.
For the parameters considered here (and indeed, for generic choices), this is an 
underdetermined linear system. We solve it using SVD and backsubstitution 
\citep{1992nrfa.book.....P}, setting
singular values $< 10^{-5}$ to zero. Fig. \ref{fig:svd_inversion} shows the estimated $dN/dz$ 
averaged over 50 simulations, and compares it to the true redshift distribution.

The first observation is that the classical solution reconstructs the redshift 
distribution accurately for certain choices of discretizations, and in particular, for 
discretizations of the photometric distributions with step sizes approximately the width of the 
photometric redshift errors. The largest errors are for $z \geq 0.9$ that result from the fact 
that $dN/dz$ is almost completely unconstrained at these redshifts 
by $[dN/dz]_{P}$ as only 6 per cent
of objects at these redshifts scatter to $z \leq 0.7$. 

We also observe that as we increase the resolution of $[dN/dz]_{P}$, the reconstruction
goes unstable, ringing at the edges of the photometric redshift catalogue. Note that the 
reconstructions in Fig. \ref{fig:svd_inversion} are averages, and the instabilities in 
a single reconstruction are significantly larger. 

This behaviour has a simple, intuitive explanation. The effect of photometric redshift errors
is to smooth away the high frequency ((redshift error)$^{-1}$)
components in $dN/dz$. However, $[dN/dz]_{P}$ has high 
frequencies due to noise in the data, and these induce large oscillations in the reconstruction.
To be more quantitative, we start with a simplified model of the photometric
errors, 
\be
\Delta (z-z_{p},z) \propto \exp\left(\frac{-(z-z_{p})^2}{2 \sigma^{2}} \right) \,\,.
\ee
A component of $dN/dz$ with frequency $k$ will be attenuated by a factor of 
\be
\int dz\, \exp\left(\frac{-(z-z_{p})^2}{2 \sigma^{2}} \right) e^{ikz} \propto 
\exp\left(-\frac{k^{2} \sigma^{2}}{2}\right) \,\,.
\label{eq:gaussianFT}
\ee
However, $[dN/dz]_{P}$ has a Poisson noise component that tends to a constant
at high frequencies. Therefore, the inversion excites high frequency modes
in the reconstruction with amplitude $\propto \exp(k^{2} \sigma^{2}/2)$. 
Eq. \ref{eq:gaussianFT} also implies that this becomes significant for modes with
$k > 1/\sigma$, agreeing with our intuitive picture. 

The effect of the discretization step size on the the stability of the 
classical solution is now clear; discretization cuts off frequencies
higher than $\sim 1/\delta z$ where $\delta z$ is the step size, filtering 
out the problematic modes. This also suggests that the ideal discretizations
have $\delta z \sim \sigma$, as demonstrated in our simulations.

\subsection{Regularization}

We would like to modify the classical solution so that it becomes less sensitive
to the inversion instability discussed in the previous section. 
In order to do so, it is useful to rephrase the classical solution 
as a minimization problem\footnote{For an alternative approach to solving
this problem, see \cite{1974AJ.....79..745L}}. 
If we define the energy functional,
\be
E_{0} = \left\arrowvert A_{ij} \left[\frac{dN}{dz}\right]_{j} - \left[\frac{dN}{dz}\right]_{P,i}
\right\arrowvert^{2} \,\,,
\label{eq:svd_energy}
\ee
then the classical solution is the value of $dN/dz$ that minimizes $E_{0}$. Given this description, 
it is trivial to include a penalty function that imposes smoothness on the reconstructed function,
\be
E = E_{0} + \lambda P \,\,,
\label{eq:reg_eq1}
\ee
where $P$ is the penalty function and $\lambda$ adjusts the relative weight of $P$ in the 
minimization of $E$\footnote{This approach appears in the literature as the method of 
regularization, the Phillips -- Twomey method, the constrained linear inversion method
and Tikhonov -- Miller regularization \citep{1992nrfa.book.....P}.}.
There are number of possible choices for the $P$ that would impose 
smoothness; we use the forward difference operator,
\be
P = \sum_{j=0}^{m-1} \left( \left[\frac{dN}{dz}\right]_{j+1} - \left[\frac{dN}{dz}\right]_{j}
\right)^{2} \,\,\, .
\label{eq:prior}
\ee

There remains the problem of choosing an appropriate value for $\lambda$. Unfortunately, 
there is no a general method for choosing an optimal value. The best that we can do 
is to define a general merit function that objectively selects an appropriate range
for $\lambda$. Based on the discussion in \cite{1986ipag.book.....C}, we use
\bea
\Xi^{2} = \frac{1}{n}\sum_{i=0}^{n-1} \left[\left(A_{ij} \left[\frac{dN}{dz}\right]_{av,j} 
- \left[\frac{dN}{dz}\right]_{P,i}
\right)\right]^2 \nonumber \\
+ \frac{1}{m}\sum_{j=0}^{m-1} \left\langle \left(\left[\frac{dN}{dz}\right]_{j} - 
\left[\frac{dN}{dz}\right]_{av,j} \right)^{2}\right\rangle \,\,,
\label{eq:merit}
\eea
where the average reconstruction $[dN/dz]_{av,j}$ is estimated either from simulations or bootstrap 
resampling.
The first term is a measure of how well the reconstructed $dN/dz$ reproduces the observed $[dN/dz]_{p}$;
this term is minimized as $\lambda \rightarrow 0$ \footnote{Assuming the generic case of an underdetermined 
system, $m \gg n$.}  and increases with increasing $\lambda$. The second term, the error in 
the reconstruction, measures its stability to the presence of noise in the data. As $\lambda \rightarrow 
\infty$, the penalty function dominates the minimization and the reconstruction is the most stable. 
As $\lambda$ decreases, the reconstruction is more sensitive to noise in the data, increasing this term. 
Choosing a value of $\lambda$ near\footnote{We are being intentionally vague here; the precise minimum may 
not be the optimal choice. However, the value of $\Xi^{2}$ provides a measure of the error that 
one is making as we move from the minimum.} the minimum of $\Xi^{2}$ picks a compromise 
between an accurate and stable reconstruction.

\begin{figure}
\begin{center}
\leavevmode
\includegraphics[width=3.0in]{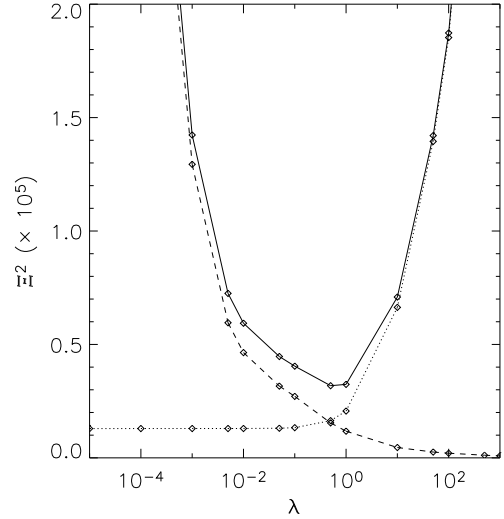}
\end{center}
\caption{The value of $\Xi^{2}$ as a function
of $\lambda$ for the simulations discussed in the text.
$\lambda$ has been rescaled such that $\lambda=1$ corresponds
to equal weight being given to $E_{0}$ and $P$ in 
Eq. \ref{eq:reg_eq1}. The dotted and dashed lines show the 
error and stability components of Eq. \ref{eq:merit} 
respectively. As expected, the error term increases with 
increasing $\lambda$, while the stability term decreases
with increasing $\lambda$. The minimum of $\Xi^{2}$ 
occurs near $\lambda = 0.5$. 
}
\label{fig:reg_inversion_xi}
\end{figure}

In order to test this method, we return to the simulations of the
previous section. Since the regularization removes the sensitivity
to the discretization of the photometric distribution, we discretize
$[dN/dz]_{P}$ into 50 bins of thickness $\Delta z=0.012$. The 
estimated redshift distribution is parametrized by  40 step functions 
of width $\Delta z=0.025$. Given these parameters, we must estimate
the appropriate value of $\lambda$. To do this, we run
50 simulations for a given value of $\lambda$ to evaluate $\Xi^{2}$ and
repeat this for a grid of values of $\lambda$. The
results are shown in Fig. \ref{fig:reg_inversion_xi}. We note that 
$\Xi^{2}$ has a well defined minimum, with the error and stability 
terms
demonstrating the $\lambda$ dependence that we anticipated. 
Note that the error term does not go to zero as $\lambda \rightarrow
0$, but appears to asymptote to a non-zero constant. This 
is readily understood in terms of the discussion in the previous section~:
the measured $[dN/dz]_{P}$ has a high-frequency noise component that 
cannot be reproduced by the convolution of $dN/dz$ with the redshift 
errors. It is this noise component that is responsible for the 
non-zero value of the error term in $\Xi^{2}$ as $\lambda \rightarrow 0$.

\begin{figure}
\begin{center}
\leavevmode
\includegraphics[width=3.0in]{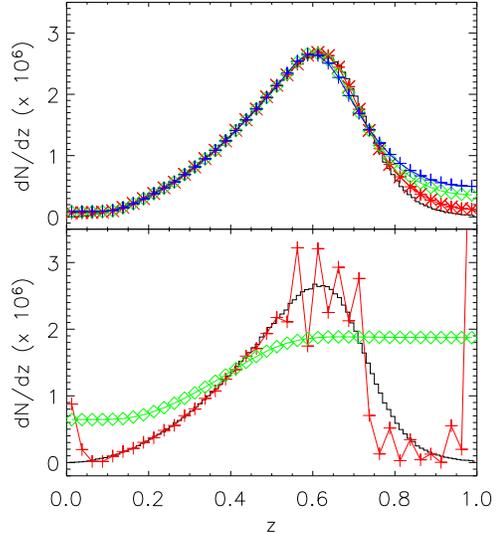}
\end{center}
\caption{Regularized estimates of $dN/dz$ for different
values of the regularization parameter, $\lambda$. In both
panels, the black histogram shows the true redshift distribution.
The upper panel shows the reconstructions for values
of $\lambda$ about the minimum of $\Xi^{2}$; the stars (red), 
diamonds (green), and crosses (blue) correspond to 
$\lambda=0.1, 0.5,$ and $1.0$ respectively. The lower panel shows 
the reconstructions for extreme values of $\lambda$, the crosses
(red) and diamonds (green) correspond to $\lambda= 10^{-10}$ and 
$1000$ respectively. The values of $\lambda$
have been rescaled as in Fig. \ref{fig:reg_inversion_xi}.
}
\label{fig:reg_inversion}
\end{figure}

The upper panel of Fig. \ref{fig:reg_inversion} shows the 
average of 50 reconstructions
of $dN/dz$ for values of $\lambda$ near the minimum of $\Xi^{2}$. We 
observe that for all the values of $\lambda$ considered in the figure,
the reconstructions closely match the input redshift distribution for
all redshifts $< 0.7$. As before, the largest discrepancies are at 
high redshift because of the lack of constraint due to the upper
photometric redshift limit of $0.7$. It is also instructive to consider
extreme values of $\lambda$; these are shown in the lower panel
of Fig.\ref{fig:reg_inversion}. For small values of $\lambda$, the 
reconstructions are extremely noisy, while for large values of $\lambda$,
the penalty function dominates the reconstruction. Note that the
forward difference operator (Eq. \ref{eq:prior}) represents a constant 
prior, which is what we see the reconstructions approaching as 
$\lambda \rightarrow \infty$.

We make one cautionary observation. Based on Fig. \ref{fig:reg_inversion},
one might conclude that the best strategy for choosing $\lambda$ is to 
preferentially choose a smaller value than what is suggested
by the minimum of $\Xi^{2}$. We however discourage this because,
as indicated in Fig.\ref{fig:reg_inversion_xi}, such reconstructions 
are very noisy. This lack of stability would result in small errors in the 
redshift error distribution  being amplified in the reconstructions. 

How many galaxies are required for the inversion? The simulations discussed above
used 100,000 galaxies, similar to the expected number of photometric LRGs over
the same area of sky. We have however tested the inversion on as few as 1000 galaxies,
and found that, for appropriate regularizations, the algorithm 
reconstructs the input redshift distribution. However, for small samples, 
the Poisson noise in the input photometric redshift distribution can be significant,
resulting in a noisier reconstruction (for the same redshift resolution). 
This may be improved by smoothing the resulting reconstruction or equivalently,
reconstructing the redshift distribution on a coarser redshift grid.

There is an important generalization of this method that should be mentioned. 
We introduced the concept of regularization and the penalty function to 
cure an instability in the deconvolution as we attempted a finer resolution 
of the redshift distribution. Phrased differently, the deconvolution became 
unstable when when the input became low S/N and the prior (in the form of the 
penalty function) compensated for this loss of information. In the cases considered
in this paper, we have used a relatively weak prior; however, if one has reliable 
prior information (for eg. the rough shape of the redshift distribution), one 
can easily include that information. A strong prior will allow one to obtain
a solution even in the low S/N regime. We do however remind the overzealous 
reader that the usual caveat about strong priors {\it does} apply in the case; the method
cannot distinguish an incorrect prior, and will get the wrong answer if such 
a prior is heavily weighted.

\subsection{Application to SDSS Data}

\begin{figure}
\begin{center}
\leavevmode
\includegraphics[width=3.0in]{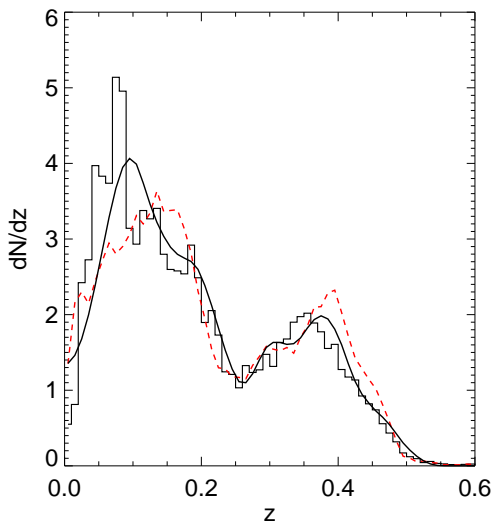}
\end{center}
\caption{Regularized estimate [solid, black] of $dN/dz$ for
the Cut I calibration data [histogram]. The input photometric 
redshift distribution is the dashed [red] line.
$dN/dz$ is normalized to integrate to unity.}
\label{fig:spectroinv}
\end{figure}

Before applying this algorithm to a photometric sample, we test it against
the Cut~I calibration dataset described in Sec.\ref{sec:select}. The results
are in Fig.\ref{fig:spectroinv}. The reconstructed redshift distribution
correctly captures all the broad features of the true redshift distribution,
including correcting for the bias at $z \sim 0.1$ and sharpening the dip
at $z \sim 0.25$. Fig.\ref{fig:spectroinv} also highlights the inability
of this method to reconstruct sharp features since these are disfavoured 
by the smoothness prior we impose; the inversion works
best for broad features. It is worth emphasizing that 
most sharp features (including the feature at $z \sim 0.075$) are spurious
(eg. binning artifacts). However, if a sharp feature is physically
expected in the distribution, the prior must be adjusted to allow 
for this.

\begin{figure}
\begin{center}
\leavevmode
\includegraphics[width=3.0in]{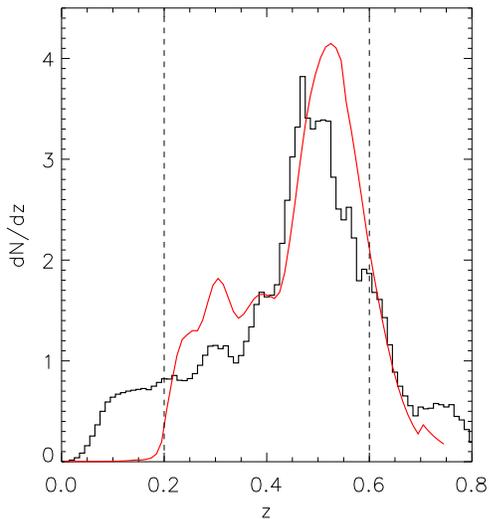}
\end{center}
\caption{Regularized estimate [solid, red] of $dN/dz$ for 
the LRG sample culled from the SDSS photometric data, 
compared with the photometric redshift distribution [histogram, black].
The redshift distribution is for galaxies with $0.2 < z_{photo} < 0.6$,
indicated by the vertical dashed lines. As before, $dN/dz$ is normalized
to integrate to unity.
}
\label{fig:lrgzdist}
\end{figure}

We conclude this discussion by applying the above algorithm to the SDSS
photometric data. A detailed discussion of the construction and
properties of the SDSS photometric LRG sample will be presented 
elsewhere; briefly, the sample is constructed by applying the 
photometric selection criteria (Cut~I and Cut~II, see Sec.~\ref{sec:select})
to objects classified as galaxies by the photometric pipeline. 
We then estimate a photometric redshift for each of the selected objects
using the simple template fitting code of Sec.~\ref{sec:single}; however, the results
are insensitive to the choice of algorithm. The 
photometric redshift distribution is shown in Fig. \ref{fig:lrgzdist}.

One feature of this distribution that deserves some explanation is the 
``bump'' in the number of galaxies at $z\sim 0.7$. This is inconsistent
with being the same population of LRGs selected with an apparent magnitude
cut. It is unlikely that these are a different population at $z\sim 0.7$, as
they would have to be a significantly brighter population than the LRGs, that only
appeared at high redshifts. A more likely 
explanation is that these are faint galaxies at lower redshifts scattered 
to high redshifts by photometric errors. This is more likely, in light of 
the fact that these galaxies have $i\sim20, g-r \sim 2$ and $r-i \sim 1$,
giving them $r \sim 21$ and $g\sim 23$. This is at the very edge (or beyond)
the photometric completeness of SDSS, and the measurements will have 
significant photometric errors ($\sim$ tenths of a magnitude). Given such photometric
errors, it is likely for the more numerous low redshift galaxies to be
scattered into the LRG colour space. 
Furthermore, the spectral templates that we use are not well constrained 
by observations for redshifts $> 0.7$.
To avoid the complications of correcting
for such contamination, we restrict our catalog to $z_{photo} < 0.6$. Similarly, as
discussed earlier, the photometric selection breaks down at low redshifts, and 
so, we impose a lower redshift cutoff of $z_{photo} > 0.2$. Note that this lower
redshift cut is imposed only to select a uniform sample; the inversion must be (and is) 
performed at all redshifts. However, the small photometric redshift error at these
redshifts minimizes the contamination from these galaxies, effectively
truncating the inverted distribution at $z \sim 0.2$.

We can now apply the inversion algorithm to estimate the true redshift
distribution, using the error distributions measured in 
Sec. \ref{sec:results}. The merit function, Eq.\ref{eq:merit}, is computed by bootstrap 
resampling the actual catalog; the measured $\Xi^{2}$ has a form similar to
Fig. \ref{fig:reg_inversion_xi}. Using the value of the regularization parameter,
$\lambda$, obtained from $\Xi^{2}$, we show the estimated redshift distribution
for galaxies with $0.2 < z_{photo} < 0.6$ in Fig. \ref{fig:lrgzdist}. 
The underestimation of the photometric redshifts at high redshifts is 
immediately apparent from the comparison of the two distributions. The 
bumps from $z=0.3$ to $0.4$ are a residual artifact of the inversion. 
These vanish for higher values of $\lambda$,
and become stronger for lower regularizations, but are 
more unstable. The value of $\lambda$ used is a balance between 
this stability and accuracy, as intended. 

\section{Discussion}

As we discussed in the Introduction, constructing a photometric redshift
catalogue involves three steps -- photometrically selecting a sample for 
which accurate photometric redshifts can be obtained, measuring
the photometric redshift error distribution for the resulting sample, and
estimating the true redshift distribution. This paper
describes all stages of this process.
\begin{itemize}
\item We describe the selection of a sample of Luminous
Red Galaxies (LRGs) using the SDSS photometric sample. These galaxies
are typically old elliptical systems with strong 4000~\AA breaks
in their continua. The shifting of this feature through the SDSS
filters make accurate photometric redshifts possible.
\item We measure the error distribution of this sample by comparing 
photometric and spectroscopic redshifts for a calibration subsample
of galaxies culled from the SDSS and SDSS-2dF spectroscopic catalogues. 
The scatter in the redshifts is approximately $\sigma \sim 0.03$ at 
redshifts less than 0.55, and increases at higher redshifts due
to increased photometric errors and uncertainties in the templates. 
\item The accuracy of the photometric redshifts is similar for
the two algorithms we consider, a simple template fit and a hybrid
algorithm that adjusts the template to better fit the observed 
colour distribution.
\end{itemize}
We have specifically used the SDSS photometric sample 
throughout this paper, both as a  example and for
its intrinsic interest. However, we emphasize that the entire 
process that we describe can be reconstructed for any multi-colour
imaging survey with appropriate filters.

Using such a photometric redshift sample requires knowing the 
conditional probability that a galaxy with a photometric redshift, 
$z_{photo}$ has a true redshift, $z_{spectro}$. Given the redshift
error distribution, this conditional probability can be readily 
estimated using Bayes' theorem if the true 
underlying redshift distribution is known. Using the fact that 
the photometric redshift distribution is the true redshift distribution 
convolved with redshift errors, we have presented a method to 
deconvolve the errors to estimate the redshift distribution. 
This method is ill-conditioned, and therefore, we use a prior 
on the smoothness of the redshift distribution to regularize 
the deconvolution. We have calibrated the relative weight of this prior
by measuring the accuracy and stability of the recovered 
redshift distributions, and we proposed a general merit function that
objectively determines this weight. 

We conclude with a few comments about this algorithm.
\begin{itemize}
\item The particulars of the sample selection are encoded 
into the photometric error distribution.
The method is therefore completely general, and applicable to 
any combination of colour selections and photometric redshift
cuts. 
\item The accuracy of the recovered redshift distribution
is determined by the accuracy of the input error distributions.
Therefore, it is {\it essential} that the calibration data used
to measure the error distribution correspond as closely as
possible to the actual data, both in photometric depth and accuracy.
One can attempt to extrapolate these distributions to fainter magnitudes
or measure them from simulations,
but with the caveat that the actual distributions may be very different
from these, and that the reconstruction could potentially be sensitive
to these errors. We emphasize that this limitation is not unique
to this methods, but affects all analyses that use photometric 
redshifts. 
\item The deconvolution algorithm is formally applicable to 
arbitrary error distributions. However, for complex error distributions
(eg. multiply peaked distributions), multiple solutions may exist
and there is no guarantee that the method will converge to the correct 
solution. This problem is avoided here by the use of photometric
pre-selection; in general, it could also be prevented by the use
of priors in the photometric redshift estimation. We {\it strongly}
recommend using one of these methods to break photometric redshift
degeneracies.
\item An advantage to this method is that the calibration data
need not sample the redshift range of interest in the same manner
as the photometric data. It suffices that it samples the 
entire range well enough to measure the error distributions. This 
allows the use of calibration sets from heterogeneous samples,
as was done in this paper.
\item The inversion algorithm presented in this paper presents an
alternative to iterative deconvolution algorithms \citep{1974AJ.....79..745L,
2003astro.ph.10038B}. As emphasized by \cite{1974AJ.....79..745L}, the two 
methods have very different mathematical philosophies; iterative methods
treat the problem as one in statistical estimation, while the philosophy
in this paper derives from the theory of integral equations. However, in 
the high S/N regime, both methods will produce similar results, and
there is little to distinguish the two. For low S/N, the deconvolution 
problem may not possess a solution, and iterative methods may not converge. 
In these cases, the algorithm presented in this paper transparently 
allows the inclusion of external information as part of the penalty
function to yield a meaningful solution. 
In cases where one possesses reliable prior information, one 
can then refine that information to yield a better solution.
\end{itemize}

N.P. acknowledges useful discussions with Michael Blanton, Chris Hirata, 
Doug Finkbeiner, Jim Gunn, David Hogg and \v{Z}eljko Ivezi\'{c}
on photometric redshift estimation techniques and regularization
methods. We thank the referee for a careful reading and suggesting
a number of improvements, from which the paper has benefited.

Funding for the creation and distribution of the SDSS Archive has 
been provided by the Alfred P. Sloan Foundation, the Participating Institutions, 
the National Aeronautics and Space Administration, the National Science Foundation,
 the U.S. Department of Energy, the Japanese Monbukagakusho, and the Max Planck Society. 
The SDSS Web site is \texttt{http://www.sdss.org/}. 

The SDSS is managed by the Astrophysical Research Consortium (ARC) for 
the Participating Institutions. The Participating Institutions are The 
University of Chicago, Fermilab, the Institute for Advanced Study, the 
Japan Participation Group, The Johns Hopkins University, Los Alamos National 
Laboratory, the Max-Planck-Institute for Astronomy (MPIA), the Max-Planck-Institute 
for Astrophysics (MPA), New Mexico State University, University of Pittsburgh, 
Princeton University, the United States Naval Observatory, and the University of Washington.

The SDSS-2dF Redshift Survey was made possible through the dedicated efforts of the
staff at the Anglo-Australian Observatory, both in creating the 2dF
instrument and supporting it on the telescope.

\bibliography{biblio,preprints}   

\begin{thebibliography}{}

\bibitem[\protect\citeauthoryear{{Abazajian} et~al.}{{Abazajian}
  et~al.}{2004}]{2004AJ....128..502A}
{Abazajian} K. et~al., 2004, \aj, 128, 502

\bibitem[\protect\citeauthoryear{{Abazajian} et~al.}{{Abazajian}
  et~al.}{2003}]{2003AJ....126.2081A}
{Abazajian} K. et~al., 2003, \aj, 126, 2081

\bibitem[\protect\citeauthoryear{{Ben{\'{\i}}tez}}{{Ben{\'{\i}}tez}}{2000}]{20%
00ApJ...536..571B}
{Ben{\'{\i}}tez} N., 2000, \apj, 536, 571

\bibitem[\protect\citeauthoryear{{Blanton} et~al.}{{Blanton}
  et~al.}{2003}]{2003AJ....125.2276B}
{Blanton} M.~R., {Lin} H., {Lupton} R.~H., {Maley} F.~M., {Young} N., {Zehavi}
  I.,  {Loveday} J., 2003, \aj, 125, 2276

\bibitem[\protect\citeauthoryear{{Bolzonella}, {Miralles}, \& {Pell{\'
  o}}}{{Bolzonella} et~al.}{2000}]{2000A&A...363..476B}
{Bolzonella} M., {Miralles} J.-M.,  {Pell{\' o}} R., 2000, \aap, 363, 476

\bibitem[\protect\citeauthoryear{{Brodwin} et~al.}{{Brodwin}
  et~al.}{2003}]{2003astro.ph.10038B}
{Brodwin} M., {Lilly} S.~J., {Porciani} C., {McCracken} H.~J., {Fevre} O.~L.,
  {Foucaud} S., {Crampton} D.,  {Mellier} Y., 2003, astro-ph/0310038

\bibitem[\protect\citeauthoryear{{Brunner}, {Connolly}, \& {Szalay}}{{Brunner}
  et~al.}{1999}]{1999ApJ...516..563B}
{Brunner} R.~J., {Connolly} A.~J.,  {Szalay} A.~S., 1999, \apj, 516, 563

\bibitem[\protect\citeauthoryear{{Bruzual} \& {Charlot}}{{Bruzual} \&
  {Charlot}}{2003}]{2003MNRAS.344.1000B}
{Bruzual} G.,  {Charlot} S., 2003, \mnras, 344, 1000

\bibitem[\protect\citeauthoryear{{Budav{\' a}ri} et~al.}{{Budav{\' a}ri}
  et~al.}{2003}]{2003ApJ...595...59B}
{Budav{\' a}ri} T. et~al., 2003, \apj, 595, 59

\bibitem[\protect\citeauthoryear{{Budav{\' a}ri} et~al.}{{Budav{\' a}ri}
  et~al.}{2000}]{2000AJ....120.1588B}
{Budav{\' a}ri} T., {Szalay} A.~S., {Connolly} A.~J., {Csabai} I.,  {Dickinson}
  M., 2000, \aj, 120, 1588

\bibitem[\protect\citeauthoryear{{Budav{\' a}ri} et~al.}{{Budav{\' a}ri}
  et~al.}{2001}]{2001AJ....121.3266B}
{Budav{\' a}ri} T., {Szalay} A.~S., {Csabai} I., {Connolly} A.~J.,  {Tsvetanov}
  Z., 2001, \aj, 121, 3266

\bibitem[\protect\citeauthoryear{{Burles} \& {Schlegel}}{{Burles} \&
  {Schlegel}}{2004}]{schlegel}
{Burles} S.,  {Schlegel} D.~J., 2004, in preparation

\bibitem[\protect\citeauthoryear{{Cannon} et~al.}{{Cannon}
  et~al.}{2003}]{cannon_aao_sdss2df}
{Cannon} R., {Croom} S., {Pimbblet} K.,  {for the SDSS-2dF Team} , 2003, AAO
  Newsletter, 103, 8

\bibitem[\protect\citeauthoryear{{Coleman}, {Wu}, \& {Weedman}}{{Coleman}
  et~al.}{1980}]{1980ApJS...43..393C}
{Coleman} G.~D., {Wu} C.-C.,  {Weedman} D.~W., 1980, \apjs, 43, 393

\bibitem[\protect\citeauthoryear{{Collister} \& {Lahav}}{{Collister} \&
  {Lahav}}{2004}]{2004PASP..116..345C}
{Collister} A.~A.,  {Lahav} O., 2004, \pasp, 116, 345

\bibitem[\protect\citeauthoryear{{Connolly} et~al.}{{Connolly}
  et~al.}{1995}]{1995AJ....110.2655C}
{Connolly} A.~J., {Csabai} I., {Szalay} A.~S., {Koo} D.~C., {Kron} R.~G.,
  {Munn} J.~A., 1995, \aj, 110, 2655

\bibitem[\protect\citeauthoryear{{Craig} \& {Brown}}{{Craig} \&
  {Brown}}{1986}]{1986ipag.book.....C}
{Craig} I.~J.~D.,  {Brown} J.~C., 1986, {Inverse problems in astronomy: A guide
  to inversion strategies for remotely sensed data}.
\newblock Adam Hilger, Ltd., 1986, 159 p.

\bibitem[\protect\citeauthoryear{{Csabai} et~al.}{{Csabai}
  et~al.}{2003}]{2003AJ....125..580C}
{Csabai} I., {Budav{\' a}ri} T., {Connolly} A.~J.,  {et al} , 2003, \aj, 125,
  580

\bibitem[\protect\citeauthoryear{{Csabai} et~al.}{{Csabai}
  et~al.}{2000}]{2000AJ....119...69C}
{Csabai} I., {Connolly} A.~J., {Szalay} A.~S.,  {Budav{\' a}ri} T., 2000, \aj,
  119, 69

\bibitem[\protect\citeauthoryear{{Eisenstein} et~al.}{{Eisenstein}
  et~al.}{2001}]{2001AJ....122.2267E}
{Eisenstein} D.~J. et~al., 2001, \aj, 122, 2267

\bibitem[\protect\citeauthoryear{{Eisenstein} et~al.}{{Eisenstein}
  et~al.}{2003}]{2003ApJ...585..694E}
{Eisenstein} D.~J. et~al., 2003, \apj, 585, 694

\bibitem[\protect\citeauthoryear{{Firth}, {Lahav}, \& {Somerville}}{{Firth}
  et~al.}{2003}]{2003MNRAS.339.1195F}
{Firth} A.~E., {Lahav} O.,  {Somerville} R.~S., 2003, \mnras, 339, 1195

\bibitem[\protect\citeauthoryear{{Fukugita} et~al.}{{Fukugita}
  et~al.}{1996}]{1996AJ....111.1748F}
{Fukugita} M., {Ichikawa} T., {Gunn} J.~E., {Doi} M., {Shimasaku} K.,
  {Schneider} D.~P., 1996, \aj, 111, 1748

\bibitem[\protect\citeauthoryear{{Gladders} \& {Yee}}{{Gladders} \&
  {Yee}}{2000}]{2000AJ....120.2148G}
{Gladders} M.~D.,  {Yee} H.~K.~C., 2000, \aj, 120, 2148

\bibitem[\protect\citeauthoryear{{Gunn} et~al.}{{Gunn}
  et~al.}{1998}]{1998AJ....116.3040G}
{Gunn} J.~E. et~al., 1998, \aj, 116, 3040

\bibitem[\protect\citeauthoryear{{Gwyn} \& {Hartwick}}{{Gwyn} \&
  {Hartwick}}{1996}]{1996ApJ...468L..77G}
{Gwyn} S.~D.~J.,  {Hartwick} F.~D.~A., 1996, \apjl, 468, L77

\bibitem[\protect\citeauthoryear{{Hamilton}}{{Hamilton}}{1985}]{1985ApJ...297.%
.371H}
{Hamilton} D., 1985, \apj, 297, 371

\bibitem[\protect\citeauthoryear{{Hogg} et~al.}{{Hogg}
  et~al.}{1998}]{1998AJ....115.1418H}
{Hogg} D.~W., {Cohen} J.~G., {Blandford} R., {Gwyn} S.~D.~J.,  {et al} , 1998,
  \aj, 115, 1418

\bibitem[\protect\citeauthoryear{{Hogg} et~al.}{{Hogg}
  et~al.}{2001}]{2001AJ....122.2129H}
{Hogg} D.~W., {Finkbeiner} D.~P., {Schlegel} D.~J.,  {Gunn} J.~E., 2001, \aj,
  122, 2129

\bibitem[\protect\citeauthoryear{{Koo}}{{Koo}}{1985}]{1985AJ.....90..418K}
{Koo} D.~C., 1985, \aj, 90, 418

\bibitem[\protect\citeauthoryear{{Le Borgne} \& {Rocca-Volmerange}}{{Le Borgne}
  \& {Rocca-Volmerange}}{2002}]{2002A&A...386..446L}
{Le Borgne} D.,  {Rocca-Volmerange} B., 2002, \aap, 386, 446

\bibitem[\protect\citeauthoryear{{Lewis} et~al.}{{Lewis}
  et~al.}{2002}]{2002MNRAS.333..279L}
{Lewis} I.~J. et~al., 2002, \mnras, 333, 279

\bibitem[\protect\citeauthoryear{{Lucy}}{{Lucy}}{1974}]{1974AJ.....79..745L}
{Lucy} L.~B., 1974, \aj, 79, 745

\bibitem[\protect\citeauthoryear{{Lupton}}{{Lupton}}{2004}]{lupton}
{Lupton} R.~H., 2004, in preparation

\bibitem[\protect\citeauthoryear{{Nichol}}{{Nichol}}{2003}]{2003astro.ph..5041%
N}
{Nichol} R.~C., 2003, Carnegie Observatories Astrophysics Series, Vol. 3, ed.
  J. S. Mulchaey, A. Dressler, and A. Oemler (Cambridge: Cambridge Univ. Press)

\bibitem[\protect\citeauthoryear{{Oke} \& {Gunn}}{{Oke} \&
  {Gunn}}{1983}]{1983ApJ...266..713O}
{Oke} J.~B.,  {Gunn} J.~E., 1983, \apj, 266, 713

\bibitem[\protect\citeauthoryear{{Pier} et~al.}{{Pier}
  et~al.}{2003}]{2003AJ....125.1559P}
{Pier} J.~R., {Munn} J.~A., {Hindsley} R.~B., {Hennessy} G.~S., {Kent} S.~M.,
  {Lupton} R.~H.,  {Ivezi{\' c}} {\v Z}., 2003, \aj, 125, 1559

\bibitem[\protect\citeauthoryear{{Press} et~al.}{{Press}
  et~al.}{1992}]{1992nrfa.book.....P}
{Press} W.~H., {Teukolsky} S.~A., {Vetterling} W.~T.,  {Flannery} B.~P., 1992,
  {Numerical recipes in FORTRAN. The art of scientific computing}.
\newblock Cambridge University Press, 1992, 2nd ed.

\bibitem[\protect\citeauthoryear{{Sawicki}, {Lin}, \& {Yee}}{{Sawicki}
  et~al.}{1997}]{1997AJ....113....1S}
{Sawicki} M.~J., {Lin} H.,  {Yee} H.~K.~C., 1997, \aj, 113, 1

\bibitem[\protect\citeauthoryear{{Schneider}, {Gunn}, \& {Hoessel}}{{Schneider}
  et~al.}{1983}]{1983ApJ...264..337S}
{Schneider} D.~P., {Gunn} J.~E.,  {Hoessel} J.~G., 1983, \apj, 264, 337

\bibitem[\protect\citeauthoryear{{Smith} et~al.}{{Smith}
  et~al.}{2002}]{2002AJ....123.2121S}
{Smith} J.~A. et~al., 2002, \aj, 123, 2121

\bibitem[\protect\citeauthoryear{{Stoughton} et~al.}{{Stoughton}
  et~al.}{2002}]{2002AJ....123..485S}
{Stoughton} C., {Lupton} R.~H., {Bernardi} M., {Blanton} M.~R.,  {et al} ,
  2002, \aj, 123, 485

\bibitem[\protect\citeauthoryear{{Strauss} et~al.}{{Strauss}
  et~al.}{2002}]{2002AJ....124.1810S}
{Strauss} M.~A. et~al., 2002, \aj, 124, 1810

\bibitem[\protect\citeauthoryear{{Wang}, {Bahcall}, \& {Turner}}{{Wang}
  et~al.}{1998}]{1998AJ....116.2081W}
{Wang} Y., {Bahcall} N.,  {Turner} E.~L., 1998, \aj, 116, 2081

\bibitem[\protect\citeauthoryear{{Willis}, {Hewett}, \& {Warren}}{{Willis}
  et~al.}{2001}]{2001MNRAS.325.1002W}
{Willis} J.~P., {Hewett} P.~C.,  {Warren} S.~J., 2001, \mnras, 325, 1002

\bibitem[\protect\citeauthoryear{{York} et~al.}{{York}
  et~al.}{2000}]{2000AJ....120.1579Y}
{York} D.~G. et~al., 2000, \aj, 120, 1579

\end{thebibliography}
\bibliographystyle{mnras}

\end{document}